\documentclass[10pt,twocolumn,twoside]{IEEEtran}

\usepackage{color,graphics,graphicx,amsmath,amsfonts,
amssymb,times,setspace,cite,fancyhdr,epstopdf}
\usepackage[active]{srcltx}

\usepackage{hyperref}

\usepackage[english]{babel} 
\usepackage{nomencl}
\usepackage{adjustbox}
\usepackage[dvipsnames]{xcolor}
\usepackage{tikz}
\usetikzlibrary{arrows.meta,positioning,fit,calc}
\usepackage{enumerate}

\definecolor{falconNavy}{HTML}{355C7D}
\definecolor{falconNavyFill}{HTML}{E8EFF5}
\definecolor{falconBlue}{HTML}{2F6F9F}
\definecolor{falconBlueFill}{HTML}{EAF3FA}
\definecolor{falconGreen}{HTML}{2E7D6E}
\definecolor{falconGreenFill}{HTML}{EAF5F2}
\definecolor{falconAmber}{HTML}{A66A2C}
\definecolor{falconAmberFill}{HTML}{FAF2E8}
\definecolor{falconRed}{HTML}{A34A4A}
\definecolor{falconRedFill}{HTML}{F8EDEE}
\definecolor{falconGray}{HTML}{6B7280}
\definecolor{falconGrayFill}{HTML}{F3F4F6}

\usepackage{array}
\usepackage{booktabs}
\usepackage{multirow}
\usepackage{makecell}
\usepackage[utf8]{inputenc}
\usepackage{algpseudocode,algorithm,algorithmicx}
\makenomenclature
\usepackage{caption}
\usepackage{subcaption}
\usepackage{stfloats}  
\usepackage{afterpage} 
\usepackage{placeins}  

\usepackage{longtable}
\usepackage{ragged2e}

\def\BibTeX{{\rmfamily B\kern-.05em{\scshape i\kern-.025em b}\kern-.08em \TeX}}

\newcommand{\ben}{\begin{equation}}
\newcommand{\een}{\end{equation}}
\newcommand{\beq}{\begin{eqnarray}}
\newcommand{\eeq}{\end{eqnarray}}

\def\log{\mbox{log}}

\usepackage{float}
\floatstyle{ruled}

\newcommand{\R}{\mathbb{R}}

\renewcommand\floatsep{0mm}
\renewcommand\textfloatsep{2mm}

\begin{document}

\title{Feasibility-Aware Learning for Network-Constrained Unit Commitment}
\author{Hien Thanh Doan, Yeji Lim, and Hongseok Kim,~\IEEEmembership{Senior Member,~IEEE}
\thanks{Hien Thanh Doan, Yeji Lim, and Hongseok Kim are with the Department of Electronic Engineering, Sogang University, 35 Baekbeom-ro, Mapo-gu, Seoul, 04107, Korea
e-mail: hongseok@sogang.ac.kr.}}

\maketitle
\begin{abstract}
Learning-based approaches to network-constrained unit commitment (NCUC) usually still rely on an optimization solver to repair or complete their predictions, which limits the speedup they can deliver. This paper proposes FALCON-UC, a feasibility-aware graph learning framework that predicts commitment and dispatch jointly and produces final schedules without solving a mixed-integer program at inference time. It combines an LSTM--GraphSAGE predictor over a fully connected generator graph, a differentiable correction that embeds commitment-dependent dispatch bounds, power balance, and PTDF-based branch flows in training, and deterministic commitment, dispatch, and ramp-coupled reserve repairs at inference. On the 2383-bus system, FALCON-UC achieves $99.09\%$ commitment accuracy and a $0.4350\%$ mean absolute cost gap with all evaluated constraints within tolerance at $0.0562$~s per instance, a $2.27\times10^{3}$ speedup over the solver. MTLR~R-SCUC-FL is slightly more accurate but still needs an optimization solver to complete each schedule, which takes $262.2$~s per instance, $4.67\times10^{3}$ times the FALCON-UC runtime. A label-only baseline
without repair violates the balance and branch limits on every 2383-bus
test instance, and FALCON-UC also meets all evaluated constraint
tolerances on a 1354-bus system.
\end{abstract}
\IEEEpeerreviewmaketitle

\section{Introduction}
\label{sec:introduction}

Network-constrained unit commitment (NCUC) determines multi-period
generator commitment and dispatch while enforcing power balance,
spinning-reserve requirements, generation limits, ramping limits,
minimum up/down-time requirements, and transmission-security
constraints. Its binary-continuous,
temporally coupled, and network-constrained structure makes repeated
solution computationally demanding for large systems and time-sensitive
scenario analyses. Because system operators repeatedly solve similar
instances under moderate variations in demand, renewable generation,
and operating conditions, previously solved cases may be exploited to
accelerate decision-making while preserving solution quality and
feasibility.

Conventional approaches for reducing the computational burden of NCUC mainly improve the optimization procedure through decomposition, constraint screening, warm-starting, active-set identification, and problem-size reduction. More recently, learning-assisted optimization has been investigated to exploit patterns contained in previously solved NCUC instances. A learning-based framework was proposed in \cite{xavier2021learning} to predict redundant constraints, generate initial feasible solutions, and identify affine subspaces containing high-quality solutions. Historical demand profiles and commitment schedules were used in \cite{ramesh2021machine} to reduce the size of the NCUC formulation, while a subsequent feasibility layer was introduced in \cite{ramesh2023feasibility} to improve the reliability of machine-learning-assisted day-ahead decisions. This method, denoted MTLR~R-SCUC-FL, solves a reduced security-constrained unit commitment (SCUC) problem after prediction and serves as a comparator in Section~\ref{sec:numerical_experiments}. A multi-layer data-driven NCUC framework incorporating feasibility compliance was also developed in \cite{feliachi2022multi}.

Because power-system operation involves both spatial and temporal dependencies, recent studies have further investigated spatio-temporal and graph-based learning for UC and NCUC. A spatio-temporal deep-learning-assisted reduced NCUC framework was proposed in \cite{ramesh2023spatio}, using graph neural networks to capture spatial correlations in the power network and recurrent neural networks to represent temporal correlations across scheduling periods. The predicted commitment schedules and critical transmission constraints were then used to construct a reduced NCUC problem for a conventional optimizer. Grid topology was incorporated into a graph convolutional network in \cite{tang2023graph} to predict unit-commitment decisions, while a topology-guided high-quality solution-learning framework was developed in \cite{gao2025topology}. In a related UC study, graph-based initial commitment prediction was combined with large-neighborhood search in \cite{yang2025solving}.

Nevertheless, existing learning-based UC and NCUC methods still present important limitations. Many approaches employ machine learning to predict redundant constraints, identify critical transmission lines, generate warm-start solutions, fix selected commitment variables, or reduce the original formulation before solving a remaining optimization problem. Consequently, final feasibility and solution quality may still depend on an optimization-based feasibility-restoration or neighborhood-search procedure. Moreover, many learning-based NCUC methods focus primarily on commitment prediction. Accurate commitment prediction alone does not guarantee a feasible dispatch because the corresponding generation schedule should also satisfy power balance, generation limits, ramping constraints, commitment--dispatch coupling, and transmission-security constraints. Even when aggregate generation closely matches total demand, an inaccurate spatial distribution of generator outputs may produce substantial branch-flow violations.

Related learning-based optimal power flow studies have shown that constraint-aware prediction and feasibility-enhancement mechanisms can improve the physical consistency of neural-network (NN) solutions \cite{pan2021deepopf,pan2020deepopf,zhao2020deepopf+,pan2022deepopf}. Constraint-enforcing learning has also been investigated in the broader optimization literature \cite{cristian2023end}. Extending these concepts to NCUC, however, is more challenging because NCUC simultaneously involves binary commitment decisions, continuous dispatch variables, temporal coupling, startup and shutdown logic, and network-security constraints.

These limitations motivate FALCON-UC (feasibility-aware learning for
commitment optimization using networks), which jointly predicts commitment
probabilities and raw dispatch. Feasibility is incorporated through
differentiable training-time dispatch correction and deterministic
inference-time commitment and dispatch repair, producing final schedules
without re-solving the complete NCUC mixed-integer program (MIP). The main
contributions are:
\begin{itemize}
\item We propose a joint commitment--dispatch predictor that combines
temporal system encoding with permutation-equivariant message passing over
a fully connected generator graph.

\item A differentiable training-correction path couples probabilistic
commitment and dispatch predictions and applies commitment-dependent
dispatch bounds, power balance, and PTDF-based branch-flow information
during training, so that the losses are computed on dispatch that satisfies
the bounds and is closer to balanced and branch-feasible.

\item Deterministic commitment, spinning-reserve, and dispatch repairs
improve generation and reserve adequacy and temporal consistency while
reducing dispatch-bound, power-balance, ramping, and branch-flow
violations without requiring an additional NCUC solve at inference time.

\item We benchmark FALCON-UC on the 2383-bus system against an unrepaired
label-only baseline, MTLR~R-SCUC-FL, and a NCUC solver, and on the 1354-bus
system against the solver. On the 2383-bus system, all constraint
violations remain within tolerance, with speedups of $4.67\times10^{3}$
over MTLR~R-SCUC-FL and $2.27\times10^{3}$ over the solver, and a maximum
cost gap ($2.56\%$) under half that of MTLR~R-SCUC-FL ($5.61\%$). The
unrepaired baseline is fully feasible on no test instance and has a
$3.79\%$ mean cost gap, versus $0.44\%$ for FALCON-UC, reflecting training
and repair together.
\end{itemize}

\section{Problem Formulation}
\label{sec:problem_formulation}

This section defines the reference NCUC model and the solution mapping approximated by FALCON-UC.

\subsection{Network-Constrained Unit Commitment Model}
\label{subsec:NCUC_model}

Consider a power system with a set of buses \(\mathcal{B}\), thermal generators \(\mathcal{G}\), transmission lines \(\mathcal{L}\), and scheduling periods \(\mathcal{T}=\{1,\ldots,T\}\). Let \(u_{g,t}\in\{0,1\}\) denote the commitment status of generator \(g\) at time \(t\), where \(u_{g,t}=1\) indicates that the unit is online. Throughout this paper, bold symbols denote vectors or matrices, whereas
nonbold symbols denote scalar quantities or individual elements.

Let \(v_{g,t}\in\{0,1\}\) and \(w_{g,t}\in\{0,1\}\) denote startup and shutdown decisions, respectively. Let \(B=|\mathcal{B}|\), \(G=|\mathcal{G}|\), and \(L=|\mathcal{L}|\).
The physical lower and upper active-power limits of generator \(g\) at time \(t\)
are denoted by \(P_{g,t}^{\min}\) and \(P_{g,t}^{\max}\), respectively.
When these limits are time-invariant, we write \(P_g^{\min}\) and
\(P_g^{\max}\) for simplicity.

The NCUC formulation represents the dispatch above the minimum generation level by \(p_{g,t}^{+}\). The total active power output of generator \(g\) at time \(t\) is defined as
\begin{equation}
P_{g,t}
=
u_{g,t}P_{g,t}^{\min}
+
p_{g,t}^{+},
\quad \forall g\in\mathcal{G},\; t\in\mathcal{T}.
\label{eq:total_generation}
\end{equation}

The deterministic NCUC formulation used for label generation and
evaluation is the default formulation of UnitCommitment.jl~v0.4.2~\cite{santos2024unitcommitment},
solved with HiGHS to a $1\%$ relative MIP gap. It includes a hard
system-wide spinning-reserve requirement
(Section~\ref{subsec:reserve_model}); contingencies are removed for the NCUC
setting. Its network model uses the default shift-factor cutoff of $0.005$,
except for six test samples, where the reference and the MTLR~R-SCUC-FL solve
use the exact shift factors to match the evaluation PTDF. The objective is to minimize the total operating cost over the scheduling horizon:
\begin{align}
\min
\quad
&
\sum_{t\in\mathcal{T}}
\sum_{g\in\mathcal{G}}
C_{g,t}^{\min} u_{g,t}
+
\sum_{t\in\mathcal{T}}
\sum_{g\in\mathcal{G}}
C_{g,t}^{\mathrm{su}} v_{g,t}
\nonumber \\
&
+
\sum_{t\in\mathcal{T}}
\sum_{g\in\mathcal{G}}
C_{g,t}^{\mathrm{p}}\left(p_{g,t}^{+}\right)
+
\sum_{t\in\mathcal{T}}
\sum_{\ell\in\mathcal{L}}
C_{\ell,t}^{\mathrm{ov}} o_{\ell,t}.
\label{eq:NCUC_objective}
\end{align}
Here, \(C_{g,t}^{\min}\) is the no-load or minimum-output cost, \(C_{g,t}^{\mathrm{su}}\) is the startup cost, \(C_{g,t}^{\mathrm{p}}(\cdot)\) is the production cost above the minimum output, \(C_{\ell,t}^{\mathrm{ov}}\) is the overload penalty for line \(\ell\), and \(o_{\ell,t}\) is the branch-flow overflow variable.

If a piecewise-linear production cost is used, let \(\mathcal{K}_g\) denote a set of production-cost segments of generator \(g\), let \(m\in\mathcal{K}_g\) denote a production-cost-segment index, let \(p_{g,t,m}^{\mathrm{seg}}\) denote the output assigned to segment \(m\), let \(\overline{P}_{g,t,m}^{\mathrm{seg}}\) denote its capacity, and let \(c_{g,t,m}^{\mathrm{seg}}\) denote its marginal cost. Then, \(p_{g,t}^{+}\) is decomposed as
\begin{equation}
p_{g,t}^{+}
=
\sum_{m\in\mathcal{K}_g}
p_{g,t,m}^{\mathrm{seg}},
\quad 0\leq p_{g,t,m}^{\mathrm{seg}}\leq\overline{P}_{g,t,m}^{\mathrm{seg}},
\quad \forall g,t,m,
\label{eq:segment_decomposition}
\end{equation}
The corresponding variable production cost is
\begin{equation}
C_{g,t}^{\mathrm{p}}\left(p_{g,t}^{+}\right)
=
\sum_{m\in\mathcal{K}_g}
c_{g,t,m}^{\mathrm{seg}}
p_{g,t,m}^{\mathrm{seg}}.
\label{eq:pwl_cost}
\end{equation}

\subsection{Commitment Logic and Generation Constraints}
\label{subsec:commitment_constraints}

Let \(u_{g,0}\) denote the given pre-horizon commitment status of generator \(g\). The commitment status, startup decision, and shutdown decision are linked, and a unit cannot start up and shut down in the same period:
\begin{equation}
u_{g,t} - u_{g,t-1} = v_{g,t} - w_{g,t},
\quad v_{g,t} + w_{g,t} \leq 1,
\quad \forall g\in\mathcal{G},\; t\in\mathcal{T}.
\label{eq:binary_link}
\end{equation}

The dispatch above the minimum output is limited by the online capacity range:
\begin{equation}
0
\leq
p_{g,t}^{+}
\leq
\left(
P_{g,t}^{\max}
-
P_{g,t}^{\min}
\right)
u_{g,t},
\quad \forall g\in\mathcal{G},\; t\in\mathcal{T},
\label{eq:prod_above_limit}
\end{equation}

Ramping constraints limit changes in total dispatch between consecutive time periods:
\begin{equation}
P_{g,t}
-
P_{g,t-1}
\leq
RU_gu_{g,t-1}
+
SU_g v_{g,t},
\quad \forall g\in\mathcal{G},\; t=2,\ldots,T,
\label{eq:ramp_up}
\end{equation}
\begin{equation}
P_{g,t-1}
-
P_{g,t}
\leq
RD_gu_{g,t}
+
SD_g w_{g,t},
\quad \forall g\in\mathcal{G},\; t=2,\ldots,T.
\label{eq:ramp_down}
\end{equation}
Here, \(RU_g\) and \(RD_g\) are the normal ramp-up and ramp-down limits, while \(SU_g\) and \(SD_g\) are the startup and shutdown ramping limits, respectively.

Minimum up-time and down-time constraints are enforced within the scheduling horizon as
\begin{equation}
\sum_{\tau=\max\{1,t-T_g^{\mathrm{up}}+1\}}^{t}
v_{g,\tau}
\leq
u_{g,t},
\quad \forall g\in\mathcal{G},\; t\in\mathcal{T},
\label{eq:min_up}
\end{equation}
\begin{equation}
\sum_{\tau=\max\{1,t-T_g^{\mathrm{down}}+1\}}^{t}
w_{g,\tau}
\leq
1-u_{g,t},
\quad \forall g\in\mathcal{G},\; t\in\mathcal{T},
\label{eq:min_down}
\end{equation}
where \(T_g^{\mathrm{up}}\) and \(T_g^{\mathrm{down}}\) denote the minimum up-time and minimum down-time of generator \(g\), respectively. The pre-horizon commitment status and the remaining minimum up/down-time obligations at the beginning of the horizon must be included in the initial-condition data.
\begin{figure*}[t]
\centering
\resizebox{0.8\textwidth}{!}{%
\begin{tikzpicture}[
    font=\sffamily\scriptsize,
    >=Latex,
    flow/.style={-Latex, line width=0.8pt, draw=falconGray},
    trainflow/.style={-Latex, line width=0.9pt, draw=falconBlue},
    testflow/.style={-Latex, line width=0.9pt, draw=falconGreen},
    backflow/.style={-Latex, dashed, line width=0.9pt, draw=falconBlue},
    box/.style={draw=falconGray, fill=white, rounded corners=1.5pt,
                line width=0.7pt, minimum height=1.15cm,
                align=center, inner sep=4pt},
    pred/.style={box, draw=falconNavy, fill=falconNavyFill},
    train/.style={box, draw=falconBlue, fill=falconBlueFill},
    test/.style={box, draw=falconGreen, fill=falconGreenFill},
    loss/.style={box, draw=falconAmber, fill=falconAmberFill},
    neutral/.style={box, draw=falconGray, fill=falconGrayFill},
    warning/.style={box, draw=falconRed, fill=falconRedFill},
    lane/.style={font=\sffamily\scriptsize\bfseries, anchor=west}
]

\node[lane, text=falconNavy] at (0.45,0.85)
    {(a) Shared predictor};
\node[pred, text width=3.2cm] (inputs) at (2.0,0)
    {\textbf{Learned inputs}\\
     System sequence $\boldsymbol{X}^{\mathrm{sys}}$\\
     Generator features $\boldsymbol{X}^{\mathrm{gen}}$};
\node[pred, text width=3.4cm] (model) at (6.2,0)
    {\textbf{Temporal--generator predictor}\\
     Temporal and generator encoders};
\node[pred, text width=3.25cm] (outputs) at (11.35,0)
    {\textbf{Predicted outputs}\\
     Commitment probabilities $\hat{\boldsymbol{u}}$\\
     Raw dispatch $\hat{\boldsymbol{P}}^{\mathrm{raw}}$};
\node[warning, text width=3.3cm, line width=1.1pt]
    (labelonly) at (16.15,0)
    {{\large\textbf{!}}\quad\textbf{Baseline endpoint}\\
     Accuracy alone does not ensure feasibility};

\draw[flow] (inputs) -- (model);
\draw[flow] (model) -- (outputs);
\draw[-Latex, dashed, line width=0.7pt, draw=falconRed]
    (outputs.east) -- (labelonly.west);

\node[lane, text=falconBlue] at (0.45,-1.30)
    {(b) Training procedure};
\node[neutral, text width=3.0cm] (auxtrain) at (2.0,-2.20)
    {\textbf{Physical data $\boldsymbol{X}_{\mathrm{aux}}$}\\
     Demand, PTDF, limits, ramps, initial states};
\node[train, text width=5.0cm] (trainrepair) at (6.85,-2.20)
    {\textbf{Differentiable dispatch correction}\\
     $\boldsymbol{P}^{(0)}
       =\hat{\boldsymbol{u}}\odot\hat{\boldsymbol{P}}^{\mathrm{raw}}$\\
     Balance $\rightarrow$ PTDF $\rightarrow$ box projection
     $(\times3)$};
\node[train, text width=2.25cm] (ptr) at (11.4,-2.20)
    {\textbf{Training output}\\
     $\boldsymbol{P}^{\mathrm{tr}}$};
\node[loss, text width=3.65cm] (losses) at (16.05,-2.20)
    {\textbf{Training objective}\\
     $\mathcal{L}_{u}+\mathcal{L}_{P}
       +\mathcal{L}_{\mathrm{cost}}$\\
     $+\mathcal{L}_{\mathrm{bal}}+\mathcal{L}_{\mathrm{br}}$};

\draw[trainflow] (auxtrain) -- (trainrepair);
\draw[trainflow] (trainrepair) -- (ptr);
\draw[trainflow] (ptr) -- (losses);
\draw[trainflow]
    (outputs.south) -- (11.35,-1.25)
    -- node[above, font=\sffamily\tiny, text=falconBlue]
       {soft predictions}
       (6.85,-1.25)
    -- (trainrepair.north);
\draw[backflow]
    (losses.south) -- (16.05,-3.05)
    -- node[pos=0.45, below, font=\sffamily\tiny,
            text=falconBlue]
       {backpropagation through correction}
       (-0.12,-3.05)
    -- (-0.12,-0.85)
    -- (4.00,-0.85)
    -- (4.00,0)
    -- (model.west);

\node[lane, text=falconGreen] at (0.45,-3.55)
    {(c) Inference procedure};
\node[test, text width=2.6cm] (threshold) at (2.0,-4.47)
    {\textbf{Hard commitment}\\
     $\hat{\boldsymbol{u}}
       \xrightarrow{\mathrm{DP}}\bar{\boldsymbol{u}}$};
\node[test, text width=3.6cm] (commitrepair) at (6.0,-4.47)
    {\textbf{Commitment repair}\\
     Min.\ up/down $\rightarrow$ capacity,\\
     ramp, branch $\rightarrow$ min.\ up/down};
\node[test, text width=2.6cm] (dispatchrepair) at (10.2,-4.47)
    {\textbf{Dispatch and reserve}\\
     $\mathcal{R}_P\rightarrow \mathcal{R}_u^{\mathrm{rsv}}$ (cycle 1)};
\node[test, text width=2.6cm] (finalout) at (13.4,-4.47)
    {\textbf{Final schedule}\\
     $(\tilde{\boldsymbol{u}},
       \tilde{\boldsymbol{P}})$};
\node[neutral, text width=2.7cm] (evaluation) at (16.85,-4.47)
    {\textbf{Evaluation}\\
     Accuracy, cost, feasibility, runtime};

\draw[testflow]
    (outputs.north) -- (11.35,1.18)
    -- (-0.54,1.18)
    -- (-0.54,-4.47)
    -- (threshold.west);
\draw[testflow] (threshold) -- (commitrepair);
\draw[testflow] (commitrepair) -- (dispatchrepair);
\draw[testflow] (dispatchrepair) -- (finalout);
\draw[testflow] (finalout) -- (evaluation);
\draw[testflow]
    (finalout.south) -- (13.4,-5.40)
    -- node[below, font=\sffamily\tiny, text=falconGreen]
              {if overload $>\epsilon_{\mathrm{ov}}$: repeat
        $\mathcal{R}_u^{\mathrm{crb}}\!\rightarrow\!\mathcal{R}_u^{\mathrm{mud}}\!\rightarrow\!\mathcal{R}_P$ once}
       (6.0,-5.40)
    -- (commitrepair.south);

\end{tikzpicture}%
}
\caption{Overview of FALCON-UC. (a) A shared temporal--generator predictor
outputs commitment probabilities and raw dispatch. (b) During training, a
differentiable dispatch correction precedes the loss, and gradients flow
through it. (c) During inference, deterministic commitment, dispatch, and
reserve repairs in Algorithm~\ref{alg:inference_repair} produce the
final schedule.}
\label{fig:FALCON-UC_framework}
\end{figure*}
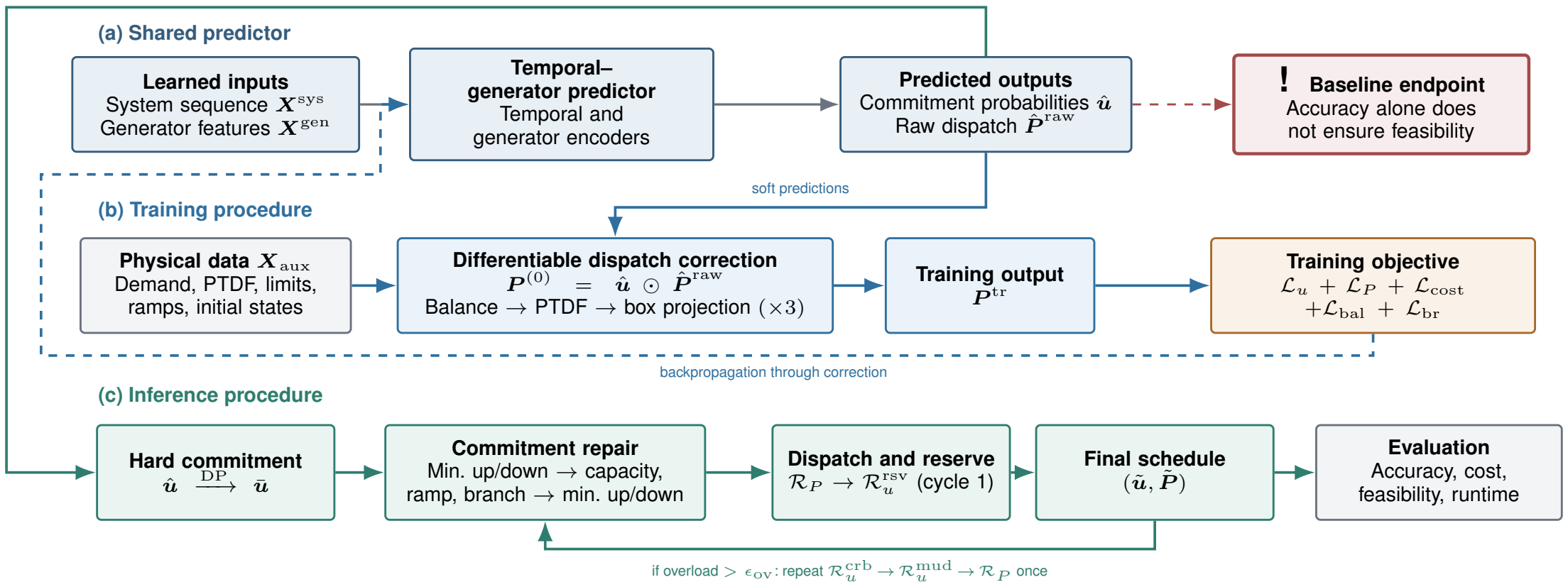

\subsection{Power Balance and Network Constraints}
\label{subsec:network_constraints}

Let \(D_{b,t}\) denote the active power demand at bus \(b\) and time \(t\), and \(\mathcal{G}_b\) be a set of generators connected to bus \(b\). The net injection at bus \(b\) is
\begin{equation}
P_{b,t}^{\mathrm{inj}}
=
\sum_{g\in\mathcal{G}_b}
P_{g,t}
-
D_{b,t},
\quad \forall b\in\mathcal{B},\; t\in\mathcal{T}.
\label{eq:net_injection}
\end{equation}
The system-level power-balance constraint is
\begin{equation}
\sum_{b\in\mathcal{B}}
P_{b,t}^{\mathrm{inj}}
=
0,
\quad \forall t\in\mathcal{T}.
\label{eq:power_balance}
\end{equation}

Under the DC power-flow approximation, let
\begin{equation}
\boldsymbol{H}
=
\left[
H_{\ell,b}
\right]_{\ell\in\mathcal{L},\,b\in\mathcal{B}}
\in\mathbb{R}^{L\times B}
\end{equation}
denote the PTDF matrix, where \(H_{\ell,b}\) is the sensitivity of the
flow on line \(\ell\) to an injection at bus \(b\) under the adopted
slack-bus convention. The flow on transmission line \(\ell\) is
\begin{equation}
F_{\ell,t}
=
\sum_{b\in\mathcal{B}}
H_{\ell,b}
P_{b,t}^{\mathrm{inj}},
\quad \forall \ell\in\mathcal{L},\; t\in\mathcal{T}.
\label{eq:ptdf_flow}
\end{equation}
The transmission-security constraints are
\begin{equation}
-F_{\ell}^{\max}-o_{\ell,t}
\leq F_{\ell,t}
\leq F_{\ell}^{\max}+o_{\ell,t},
\quad \forall \ell\in\mathcal{L},\; t\in\mathcal{T},
\label{eq:flow_limits}
\end{equation}
where \(F_{\ell}^{\max}>0\) is the thermal limit and
\(o_{\ell,t}\geq0\) is the overflow slack.

\subsection{Spinning-Reserve Requirement}
\label{subsec:reserve_model}

The reference NCUC also enforces a hard system-wide spinning-reserve
requirement \(R^{\mathrm{req}}_t\): the reserves \(r_{g,t}\geq0\) satisfy
\begin{equation}
\sum_{g\in\mathcal{G}} r_{g,t}\geq R^{\mathrm{req}}_t,
\qquad
P_{g,t}+r_{g,t}\leq\overline{P}^{\mathrm{rsv}}_{g,t},
\quad \forall g,t,
\label{eq:rsv_req}
\end{equation}
where \(\overline{P}^{\mathrm{rsv}}_{g,t}\) is the reserve-eligible output
ceiling, so that \(\sum_g[\overline{P}^{\mathrm{rsv}}_{g,t}-P_{g,t}]_{+}\) is
the reserve headroom, which ignores the ramp rows; \(r_{g,t}\) also tightens
the ramp rows of hours \(t\) and \(t+1\). FALCON-UC is trained on the solver schedules, which already satisfy this
requirement, and it does not predict \(r_{g,t}\). It receives
\(R^{\mathrm{req}}_t\) as an input and enforces the requirement through the
inference-time repairs of Section~\ref{subsubsec:reserve_repair}. Each schedule
is then verified by summing the largest reserve every unit can deliver within
its ramp limits and comparing the total with \(R^{\mathrm{req}}_t\). Because a
unit's reserve does not enter any other unit's constraints, this check is
equivalent to the solver's reserve constraints, and all stored solver schedules
pass it. The same check is applied to all methods in
Table~\ref{tab:feasibility}, and the FALCON-UC costs include the final repair.
The remaining reserve definitions are omitted for space.

\subsection{Learning-Based NCUC Approximation}
\label{subsec:learning_approximation}

Let \(\boldsymbol{X}_{\mathrm{NN}}\) denote the NN input
features, and let \(\theta\) denote all trainable parameters of
\(f_{\theta}\). FALCON-UC learns the parametric mapping
\begin{equation}
f_{\theta}:
\boldsymbol{X}_{\mathrm{NN}}
\mapsto
\left(
\hat{\boldsymbol{u}},
\hat{\boldsymbol{P}}^{\mathrm{raw}}
\right),
\label{eq:FALCON-UC_mapping}
\end{equation}
where \(\hat{\boldsymbol{u}}\in[0,1]^{G\times T}\) contains commitment
probabilities and \(\hat{\boldsymbol{P}}^{\mathrm{raw}}\in\mathbb{R}^{G\times T}\)
is total active dispatch before correction or repair, rather than the
above-minimum variable \(p_{g,t}^{+}\). The reference labels are
\((\boldsymbol{u}^{*},\boldsymbol{P}^{*})\). Section~\ref{sec:proposed_method}
describes how training and repair match these labels while reducing violations.

\section{Proposed Method}
\label{sec:proposed_method}

Figure~\ref{fig:FALCON-UC_framework} shows the shared commitment--dispatch
predictor and its two feasibility paths: differentiable dispatch correction
during training and deterministic schedule repair for validation and testing.

\subsection{Problem Setting}

The model uses system-level temporal features describing aggregate
power-system conditions and generator-level features for individual
units. For each scenario, the system-level input forms a sequence over
the scheduling horizon, with its vector at time \(t\) defined as
\begin{equation}
\boldsymbol{x}^{\mathrm{sys}}_t
=
\left[
D_t,\,
R^{\mathrm{req}}_t,\,
t/T
\right]^{\top}
\in\mathbb{R}^{3}.
\label{eq:system_feature_vector}
\end{equation}
Here, \(D_t=\sum_{b\in\mathcal{B}}D_{b,t}\) is the total demand,
\(R^{\mathrm{req}}_t\) the reserve requirement, and \(t/T\) the normalized
hour. Collecting these vectors over all \(T\) periods gives
\(\boldsymbol{X}^{\mathrm{sys}}=[\boldsymbol{x}^{\mathrm{sys}}_1,\ldots,\boldsymbol{x}^{\mathrm{sys}}_T]\in\mathbb{R}^{3\times T}\).

Let \(N_{\mathrm{f}}\) denote the number of generator-level input features, and let
\(\boldsymbol{x}^{\mathrm{gen}}_g\in\mathbb{R}^{N_{\mathrm{f}}}\) denote the
static feature vector of generator \(g\), defined as
\begin{equation}
\begin{aligned}
\boldsymbol{x}^{\mathrm{gen}}_g
=
\bigl[
&P_g^{\min},\,
P_g^{\max},\,
T_g^{\mathrm{up}},\,
T_g^{\mathrm{down}},\,
u_{g,0},\,
C_g^{\mathrm{su}},\\
&C_g^{\mathrm{mc}},\,
RU_g,\,
RD_g,\,
SU_g,\,
SD_g
\bigr]^{\top}
\in\mathbb{R}^{N_{\mathrm{f}}}.
\end{aligned}
\label{eq:generator_feature_vector}
\end{equation}
Here, \(C_g^{\mathrm{su}}\) and \(C_g^{\mathrm{mc}}\) denote the
time-invariant startup and marginal costs used as static generator
features. The remaining generator parameters are defined in
Section~\ref{sec:problem_formulation}.

The generator-level input matrix is
\(\boldsymbol{X}^{\mathrm{gen}}=[\boldsymbol{x}^{\mathrm{gen}}_1,\ldots,\boldsymbol{x}^{\mathrm{gen}}_G]^{\top}\in\mathbb{R}^{G\times N_{\mathrm{f}}}\),
and the complete NN input is
\(\boldsymbol{X}_{\mathrm{NN}}=(\boldsymbol{X}^{\mathrm{sys}},\boldsymbol{X}^{\mathrm{gen}})\).

\subsection{Neural Architecture}

The temporal encoder processes the system-level time-series features using a long short-term memory (LSTM). It produces a time-dependent embedding
\begin{equation}
\boldsymbol{h}^{\mathrm{sys}}_t
=
\mathrm{LSTM}
\left(
\boldsymbol{x}^{\mathrm{sys}}_1,\ldots,\boldsymbol{x}^{\mathrm{sys}}_T
\right)_t,
\quad t=1,\ldots,T.
\label{eq:lstm_embedding}
\end{equation}
For the generator encoder, a multilayer perceptron (MLP) produces the
initial generator embedding
\begin{equation}
\boldsymbol{h}^{(0)}_g
=
\mathrm{MLP}_{\mathrm{gen}}
\left(
\boldsymbol{x}^{\mathrm{gen}}_g
\right).
\end{equation}
Because the generator graph is fully connected,
\(\mathcal{N}(g)=\mathcal{G}\setminus\{g\}\). Let
\(N_{\mathrm{GNN}}\) denote the number of GraphSAGE layers. For
\(r=0,\ldots,N_{\mathrm{GNN}}-1\), a GraphSAGE-style update aggregates neighboring generator representations as
\begin{equation}
\boldsymbol{m}^{(r)}_g
=
\frac{1}{|\mathcal{N}(g)|}
\sum_{j\in \mathcal{G}\setminus\{g\}}
\boldsymbol{h}^{(r)}_j.
\label{eq:graphsage_aggregation}
\end{equation}
Let \([\cdot\,;\,\cdot]\) denote vector concatenation. The generator representation is updated as
\begin{equation}
\boldsymbol{h}^{(r+1)}_g
=
\sigma_{\mathrm{act}}
\left(
\boldsymbol{W}^{(r)}
\left[
\boldsymbol{h}^{(r)}_g ;
\boldsymbol{m}^{(r)}_g
\right]
+
\boldsymbol{b}^{(r)}
\right),
\label{eq:graphsage_update}
\end{equation}
where \(\boldsymbol{W}^{(r)}\) and \(\boldsymbol{b}^{(r)}\) denote
the trainable weight matrix and bias vector of GraphSAGE layer \(r\),
respectively, and \(\sigma_{\mathrm{act}}(\cdot)\) is a nonlinear activation function.
The final generator embedding is
\begin{equation}
\boldsymbol{h}^{\mathrm{gen}}_g
=
\boldsymbol{h}^{(N_{\mathrm{GNN}})}_g.
\end{equation}

The time and generator embeddings are fused for each generator--time pair:
\begin{equation}
\boldsymbol{z}_{g,t}
=
\mathrm{MLP}_{\mathrm{fuse}}
\left(
\left[
\boldsymbol{h}^{\mathrm{gen}}_g ;
\boldsymbol{h}^{\mathrm{sys}}_t
\right]
\right).
\label{eq:fusion_embedding}
\end{equation}
Then, the commitment head predicts the commitment logit
\begin{equation}
a^u_{g,t}
=
\mathrm{MLP}_{u}
\left(
\boldsymbol{z}_{g,t}
\right),
\label{eq:commitment_logit}
\end{equation}
and the corresponding commitment probability is
\begin{equation}
\hat{u}_{g,t}
=
\mathrm{sigmoid}
\left(
a^u_{g,t}
\right).
\label{eq:commitment_probability}
\end{equation}

The dispatch head predicts a scaled dispatch value:
\begin{equation}
\hat{P}^{\mathrm{sc}}_{g,t}
=
\mathrm{sigmoid}
\left(
\mathrm{MLP}_{P}
\left(
\boldsymbol{z}_{g,t}
\right)
\right).
\label{eq:scaled_dispatch}
\end{equation}
The scaled dispatch is converted to MW using the physical generation limits:
\begin{equation}
\hat{P}^{\mathrm{raw}}_{g,t}
=
P_{g,t}^{\min}
+
\hat{P}^{\mathrm{sc}}_{g,t}
\left(
P_{g,t}^{\max}
-
P_{g,t}^{\min}
\right).
\label{eq:pg_recovery}
\end{equation}

\subsection{Feasibility-Aware Correction}
\label{subsec:repair_layer}

Let \(\boldsymbol{X}_{\mathrm{aux}}\) collect the demand, reserve requirement,
generator and network parameters, and initial-condition data used by both the
training correction and the inference repair
(Section~\ref{subsubsec:validation_testing}); this dependence is suppressed in
all operators below.

\subsubsection{Differentiable Equality Correction}

During training, the \textit{soft} commitment prediction \(\hat{\boldsymbol{u}}\) is used as the
commitment input to preserve gradient propagation through the correction procedure:
\begin{equation}
\boldsymbol{u}^{\mathrm{cor}}
=
\hat{\boldsymbol{u}}.
\label{eq:training_repair_commitment_input}
\end{equation}
The corresponding commitment-dependent dispatch bounds are
\begin{equation}
\underline{P}^{\mathrm{cor}}_{g,t}
=
P_{g,t}^{\min}
u^{\mathrm{cor}}_{g,t},
\qquad
\overline{P}^{\mathrm{cor}}_{g,t}
=
P_{g,t}^{\max}
u^{\mathrm{cor}}_{g,t}.
\label{eq:training_repair_bounds}
\end{equation}

To initialize the correction within these commitment-dependent bounds, the initial dispatch is defined as
\begin{equation}
P^{(0)}_{g,t}
=
u^{\mathrm{cor}}_{g,t}
\hat{P}^{\mathrm{raw}}_{g,t}.
\label{eq:initial_training_dispatch}
\end{equation}
The initial dispatch \(\boldsymbol{P}^{(0)}\) is processed through
\(N_{\mathrm{step}}\) iterations of balance and branch-flow correction. At iteration
\(k=0,\ldots,N_{\mathrm{step}}-1\), the period-wise power mismatch is
\begin{equation}
\Delta P_t^{(k)}
=
\sum_{g\in\mathcal{G}}P_{g,t}^{(k)}
-
\sum_{b\in\mathcal{B}}D_{b,t},
\qquad t\in\mathcal{T}.
\label{eq:training_balance_mismatch}
\end{equation}
Let \([z]_{+}=\max(z,0)\). The available upward and downward
headrooms are
\begin{equation}
M_{g,t}^{\mathrm{up},(k)}
=
\left[
\overline{P}_{g,t}^{\mathrm{cor}}-P_{g,t}^{(k)}
\right]_{+},
\qquad
M_{g,t}^{\mathrm{down},(k)}
=
\left[
P_{g,t}^{(k)}-\underline{P}_{g,t}^{\mathrm{cor}}
\right]_{+}.
\label{eq:balance_headroom}
\end{equation}
The directional headroom is selected according to the sign of the
mismatch:
\begin{equation}
M_{g,t}^{\mathrm{dir},(k)}
=
\begin{cases}
M_{g,t}^{\mathrm{up},(k)},
& \Delta P_t^{(k)}<0,\\
M_{g,t}^{\mathrm{down},(k)},
& \Delta P_t^{(k)}>0,\\
0,
& \Delta P_t^{(k)}=0.
\end{cases}
\end{equation}
The corresponding allocation weight is
\begin{equation}
\omega_{g,t}^{(k)}
=
\begin{cases}
\displaystyle
\frac{M_{g,t}^{\mathrm{dir},(k)}}
{\sum_{j\in\mathcal{G}}M_{j,t}^{\mathrm{dir},(k)}},
&
\displaystyle
\sum_{j\in\mathcal{G}}M_{j,t}^{\mathrm{dir},(k)}>0,\\[3mm]
0,
& \text{otherwise}.
\end{cases}
\label{eq:branch_repair_balance_weight}
\end{equation}
The portion of the mismatch that can be allocated within the available
headroom is
\begin{equation}
\Delta P_t^{\mathrm{alloc},(k)}
=
\operatorname{sgn}\!\left(\Delta P_t^{(k)}\right)
\min\!\left\{
\left|\Delta P_t^{(k)}\right|,
\sum_{j\in\mathcal{G}}M_{j,t}^{\mathrm{dir},(k)}
\right\}.
\end{equation}
The resulting balance-corrected dispatch defines the
headroom-weighted balance-correction operator:
\begin{equation}
\begin{aligned}
P_{g,t}^{\mathrm{bal},(k)}
&=
\left[
\mathcal{R}_P^{\mathrm{bal}}
\left(\boldsymbol{P}^{(k)}\right)
\right]_{g,t} \\
&=
P_{g,t}^{(k)}
-
\Delta P_t^{\mathrm{alloc},(k)}
\omega_{g,t}^{(k)},
\end{aligned}
\label{eq:branch_repair_balance_correction}
\end{equation}
for all \(g\in\mathcal{G}\) and \(t\in\mathcal{T}\), where the dependence
of \(\mathcal{R}_P^{\mathrm{bal}}\) on demand and the commitment-dependent bounds
is suppressed for compactness. 
The mismatch is eliminated when
sufficient aggregate directional headroom is available; otherwise, the
unallocable portion remains as a residual.

\subsubsection{Differentiable Inequality Correction} Let \(\boldsymbol{A}\in\{0,1\}^{B\times G}\) denote the
generator-to-bus incidence matrix,
\(\boldsymbol{d}_t\in\mathbb{R}^{B}\) the bus-demand vector at time
\(t\), \(\boldsymbol{e}_s\in\{0,1\}^{B}\) the unit vector associated
with slack bus \(s\), and
\(\boldsymbol{1}\in\mathbb{R}^{B}\) the \(B\)-dimensional all-ones
vector. The PTDF matrix
\(\boldsymbol{H}\in\mathbb{R}^{L\times B}\) is as defined in
Section~\ref{subsec:network_constraints}. The slack-adjusted net-injection and branch-flow vectors are
\begin{align}
\widetilde{\boldsymbol{P}}^{\mathrm{inj},(k)}_t
&=
\boldsymbol{A}
\boldsymbol{P}^{\mathrm{bal},(k)}_t
-
\boldsymbol{d}_t
-
\boldsymbol{e}_s
\boldsymbol{1}^{\top}
\left(
\boldsymbol{A}
\boldsymbol{P}^{\mathrm{bal},(k)}_t
-
\boldsymbol{d}_t
\right),
\label{eq:branch_repair_net_injection}\\
\boldsymbol{F}^{(k)}_t
&=
\boldsymbol{H}
\widetilde{\boldsymbol{P}}^{\mathrm{inj},(k)}_t.
\label{eq:branch_repair_flow}
\end{align}
For each branch \(\ell\in\mathcal{L}\), the normalized branch-flow violation is
\begin{equation}
\nu_{\ell,t}^{(k)}
=
\frac{
\left[
\left|F_{\ell,t}^{(k)}\right|-F_{\ell}^{\max}
\right]_{+}
}{
\max\!\left(F_{\ell}^{\max},\epsilon\right)
}.
\label{eq:normalized_branch_violation}
\end{equation}
where \(\epsilon>0\) is a small numerical constant. The absolute value accounts for both flow directions.

Let \(\beta_{\max}\geq 0\) control the emphasis placed on the most
severely overloaded branch, let \(\overline{C}(\cdot)\) denote a
dimensionless normalized approximation of generation cost, and let
\(\lambda_{\mathrm{c}}\geq 0\) denote its weight. The branch-correction objective is then given by
\begin{equation}
\begin{aligned}
\Phi_{\mathrm{br}}\!\left(\boldsymbol{P}^{\mathrm{bal},(k)}\right)
={}&
\frac{1}{|\mathcal{L}||\mathcal{T}|}
\sum_{t\in\mathcal{T}}
\sum_{\ell\in\mathcal{L}}
\left(\nu_{\ell,t}^{(k)}\right)^2 \\
&+
\frac{\beta_{\max}}{|\mathcal{T}|}
\sum_{t\in\mathcal{T}}
\left(
\max_{\ell\in\mathcal{L}}
\nu_{\ell,t}^{(k)}
\right)^2
+
\lambda_{\mathrm{c}}
\overline{C}\!\left(
\boldsymbol{P}^{\mathrm{bal},(k)}
\right).
\end{aligned}
\label{eq:branch_repair_objective}
\end{equation}

Our goal is to adjust \(\boldsymbol{P}^{\mathrm{bal},(k)}\) so that the flows in~\eqref{eq:branch_repair_net_injection}--\eqref{eq:branch_repair_flow} satisfy their thermal limits at low generation cost without undoing the balance correction. The gradient of the branch-correction objective is
\begin{equation}
\nabla^{(k)}_{g,t}
=
\frac{
\partial
\Phi_{\mathrm{br}}
\left(
\boldsymbol{P}^{\mathrm{bal},(k)}
\right)
}{
\partial
P^{\mathrm{bal},(k)}_{g,t}
}.
\label{eq:branch_repair_raw_gradient}
\end{equation}
The absolute-flow formulation accounts for both forward- and
reverse-direction thermal overloads. Let \(\mathbb{I}(\cdot)\) denote the indicator function and define
\begin{equation}
\chi_{g,t}
=
\mathbb{I}
\left(
\overline{P}^{\mathrm{cor}}_{g,t}
>
\underline{P}^{\mathrm{cor}}_{g,t}
\right).
\label{eq:branch_repair_adjustable_mask}
\end{equation} Thus,
\(\chi_{g,t}=1\) if generator \(g\) has an adjustable dispatch range at
time \(t\), and \(\chi_{g,t}=0\) otherwise. The mean gradient over adjustable generators is
\begin{equation}
\bar{\nabla}^{(k)}_{t}
=
\frac{
\sum_{g\in\mathcal{G}}
\chi_{g,t}\nabla^{(k)}_{g,t}
}{
\max\!\left(
1,
\sum_{g\in\mathcal{G}}
\chi_{g,t}
\right)
},
\label{eq:branch_repair_mean_gradient}
\end{equation}
and the corresponding zero-sum projected gradient is
\begin{equation}
\nabla^{\perp,(k)}_{g,t}
=
\chi_{g,t}
\left(
\nabla^{(k)}_{g,t}
-
\bar{\nabla}^{(k)}_{t}
\right).
\label{eq:branch_repair_projected_gradient}
\end{equation}

Let \(\Pi_{\mathrm{box}}(\cdot)\) denote the componentwise projection onto
these bounds:
\begin{equation}
\left[
\Pi_{\mathrm{box}}(\boldsymbol{Z})
\right]_{g,t}
=
\Pi_{\left[
\underline{P}^{\mathrm{cor}}_{g,t},
\overline{P}^{\mathrm{cor}}_{g,t}
\right]}
\left(
Z_{g,t}
\right),
\label{eq:training_box_projection}
\end{equation}
where the scalar interval projection is
\begin{equation}
\Pi_{[a,b]}(z)
=
\min\!\left\{
\max\!\left(z,a\right),
b
\right\}.
\end{equation} Using this projection, let \(\eta>0\) denote the correction step size.
The dispatch is then updated as
\begin{equation}
\boldsymbol{P}^{(k+1)}
=
\Pi_{\mathrm{box}}
\left(
\boldsymbol{P}^{\mathrm{bal},(k)}
-
\eta
\boldsymbol{\nabla}^{\perp,(k)}
\right),
\quad
k=0,\ldots,N_{\mathrm{step}}-1.
\label{eq:iterative_branch_gradient_repair}
\end{equation} The output of the differentiable training correction is denoted by
\begin{equation}
\boldsymbol{P}^{\mathrm{tr}}
=
\boldsymbol{P}^{(N_{\mathrm{step}})}.
\label{eq:output_diff_training_correction}
\end{equation}
Interestingly, we find that only a few correction, e.g., \(N_{\mathrm{step}}=3\), is enough for practice. Each training-correction iteration applies balance correction, a
zero-sum branch-gradient update, and commitment-dependent box projection, as
summarized in Fig.~\ref{fig:FALCON-UC_framework}.

\subsubsection{Training Objective}
\label{subsec:training_objective}

The training losses are computed using the differentiable
training-correction output \(\boldsymbol{P}^{\mathrm{tr}}\) in \eqref{eq:output_diff_training_correction}. Using the reference commitment \(u^*_{g,t}\) and the total-dispatch label \(P^*_{g,t}\) defined in Section~\ref{subsec:learning_approximation}, the commitment logit $a^u_{g,t}$ in \eqref{eq:commitment_logit}, and the generator weight denoted by \(\rho_g>0\), the weighted commitment loss is
\begin{equation}
\mathcal{L}_{u}
=
\frac{1}{|\mathcal{T}|\sum_{g\in\mathcal{G}}\rho_g}
\sum_{t\in\mathcal{T}}
\sum_{g\in\mathcal{G}}
\rho_g
\left[
\log\!\left(1+\exp(a^u_{g,t})\right)
-
u^*_{g,t}a^u_{g,t}
\right].
\label{eq:commitment_loss}
\end{equation}
We also have the dispatch loss,
\begin{equation}
\mathcal{L}_{P}
=
\frac{1}{\alpha_P |\mathcal{T}|\sum_{g\in\mathcal{G}}\rho_g}
\sum_{t\in\mathcal{T}}
\sum_{g\in\mathcal{G}}
\rho_g
\left(
P^{\mathrm{tr}}_{g,t}
-
P^*_{g,t}
\right)^2,
\label{eq:dispatch_loss}
\end{equation}
where \(\alpha_P>0\) is a fixed normalization constant with units of \(\mathrm{MW}^2\).

\begin{algorithm}[t]
\caption{FALCON-UC inference and feasibility repair}
\label{alg:inference_repair}
\begin{algorithmic}[1]
\Require \(\boldsymbol{X}_{\mathrm{NN}}\), \(\boldsymbol{X}_{\mathrm{aux}}\), \(\lambda_{\mathrm{su}}\)
\Ensure Repaired schedule \((\tilde{\boldsymbol{u}},\tilde{\boldsymbol{P}})\)
\State \((\hat{\boldsymbol{u}},\hat{\boldsymbol{P}}^{\mathrm{raw}})
       \gets f_{\theta}(\boldsymbol{X}_{\mathrm{NN}})\)
\State \(\bar{\boldsymbol{u}}\gets\) decode \(\hat{\boldsymbol{u}}\) by DP
\State \(\boldsymbol{u}\gets \mathcal{R}_u^{\mathrm{mud}}(\bar{\boldsymbol{u}})\);
       \(\boldsymbol{P}^{\mathrm{score}}\gets\hat{\boldsymbol{P}}^{\mathrm{raw}}\)
\For{\(c=1,2\)}
  \State \(\boldsymbol{u}\gets \mathcal{R}_u^{\mathrm{crb}}
         (\boldsymbol{u},\hat{\boldsymbol{u}},\boldsymbol{P}^{\mathrm{score}})\)
         \Comment{capacity, ramp, branch}
  \State \(\boldsymbol{u}\gets \mathcal{R}_u^{\mathrm{mud}}(\boldsymbol{u})\)
  \State \(\boldsymbol{P}\gets \mathcal{R}_P(\hat{\boldsymbol{P}}^{\mathrm{raw}},\boldsymbol{u})\)
  \If{\(c=1\)}
    \State \((\boldsymbol{u},\boldsymbol{P})\gets \mathcal{R}_u^{\mathrm{rsv}}(\boldsymbol{u},\boldsymbol{P})\)
           \Comment{Section~\ref{subsubsec:reserve_repair}}
  \EndIf
    \If{max.\ overload of \(\boldsymbol{P}\) over active branches \(\leq\epsilon_{\mathrm{ov}}\)}
    \State \textbf{break}
  \EndIf
  \State \(\boldsymbol{P}^{\mathrm{score}}\gets\boldsymbol{P}\)
\EndFor
\State \((\tilde{\boldsymbol{u}},\tilde{\boldsymbol{P}})
       \gets(\boldsymbol{u},\boldsymbol{P})\)
\State \(\tilde{\boldsymbol{P}}\gets\mathcal{R}_P^{\mathrm{rsv}}(\tilde{\boldsymbol{u}},\tilde{\boldsymbol{P}})\)
       \Comment{Section~\ref{subsubsec:reserve_repair}}
\State Certify \((\tilde{\boldsymbol{u}},\tilde{\boldsymbol{P}})\); if rejected,
       apply the fallback in Section~\ref{subsec:fallback}
\end{algorithmic}
\end{algorithm}

Let \(\mathcal{I}_{\mathrm{mb}}\) denote the minibatch sample-index
set. For each \(n\in\mathcal{I}_{\mathrm{mb}}\), let
\(\widetilde{C}^{\mathrm{tr}}_n\) and \(C^{*}_n\) denote the total
operating costs of the differentiably corrected and solver-reference
solutions, respectively. Here, total operating cost is the sum of
startup and generation costs and excludes the branch-overflow penalty.
The cost loss is
\begin{equation}
\mathcal{L}_{\mathrm{cost}}
=
\frac{1}{|\mathcal{I}_{\mathrm{mb}}|}
\sum_{n\in\mathcal{I}_{\mathrm{mb}}}
\left(
\frac{
\widetilde{C}^{\mathrm{tr}}_n - C^{*}_n
}{
\left|C^{*}_n\right| + \epsilon_C
}
\right)^2.
\label{eq:cost_loss}
\end{equation}
where \(\epsilon_C=10^{-6}\) is a small numerical constant used to
avoid division by zero.
The post-correction power-balance loss is
\begin{equation}
\mathcal{L}_{\mathrm{bal}}
=
\frac{1}{|\mathcal{T}|}
\sum_{t\in\mathcal{T}}
\left|
\sum_{g\in\mathcal{G}}
P^{\mathrm{tr}}_{g,t}
-
\sum_{b\in\mathcal{B}}
D_{b,t}
\right|.
\label{eq:balance_loss}
\end{equation}

Let \(\widetilde{\boldsymbol{F}}^{\mathrm{tr}}_t\) denote the branch flows obtained from
\eqref{eq:branch_repair_net_injection}--\eqref{eq:branch_repair_flow} with
\(\boldsymbol{P}^{\mathrm{bal},(k)}_t\) replaced by \(\boldsymbol{P}^{\mathrm{tr}}_t\).
Let
\(\boldsymbol{F}^{\max}
=
[F_{\ell}^{\max}]_{\ell\in\mathcal{L}}
\in\mathbb{R}^{L}\)
denote the vector of branch-flow limits. The post-correction branch-flow loss is:
\begin{equation}
\mathcal{L}_{\mathrm{br}}
=
\frac{1}{|\mathcal{T}|}
\sum_{t\in\mathcal{T}}
\left\|
\left[
\left|
\widetilde{\boldsymbol{F}}_{t}^{\mathrm{tr}}
\right|
-
\boldsymbol{F}^{\max}
\right]_{+}
\right\|_{1},
\label{eq:branch_flow_loss}
\end{equation}
where the absolute-value and positive-part operations are applied
componentwise.

Finally, the total training loss is
\begin{equation}
\begin{aligned}
\mathcal{L}
={}&
\lambda_u \mathcal{L}_{u}
+
\lambda_P \mathcal{L}_{P}
+
\lambda_{\mathrm{cost}} \mathcal{L}_{\mathrm{cost}}
\\
&+
\lambda_{\mathrm{bal}} \mathcal{L}_{\mathrm{bal}}
+
\lambda_{\mathrm{br}} \mathcal{L}_{\mathrm{br}},
\end{aligned}
\label{eq:total_loss}
\end{equation}
where \(\lambda_u\), \(\lambda_P\), \(\lambda_{\mathrm{cost}}\), \(\lambda_{\mathrm{bal}}\), and \(\lambda_{\mathrm{br}}\) are nonnegative hyperparameters selected by hyperparameter tuning in Section~\ref{subsec:experimental_setup}.

\subsection{Inference-Time Feasibility Repair}
\label{subsubsec:validation_testing}
 
At inference, the model's soft outputs must be turned into a binary
commitment. This is done in two steps. First, dynamic programming (DP) decodes a preliminary schedule
\(\bar{\boldsymbol{u}}_g\) for each unit over the whole horizon. The DP
maximizes the log-likelihood of \(\hat{\boldsymbol{u}}_g\) minus a fixed
penalty \(\lambda_{\mathrm{su}}\) for each startup (\(\lambda_{\mathrm{su}}=2\),
chosen on validation). It enforces minimum up/down times and the initial
status. It also forbids startups for units with \(P^{\min}_g > SU_g\) and
shutdowns for units with \(P^{\min}_g > SD_g\).

Second, Algorithm~\ref{alg:inference_repair} repairs this schedule with
five deterministic operators:
(i) a minimum up/down-time repair \(\mathcal{R}_u^{\mathrm{mud}}\);
(ii) a capacity-, ramp-, and branch-aware commitment repair
\(\mathcal{R}_u^{\mathrm{crb}}\);
(iii) a dispatch repair \(\mathcal{R}_P\) under fixed commitment;
(iv) a reserve repair \(\mathcal{R}_u^{\mathrm{rsv}}\); and
(v) a ramp-coupled reserve repair \(\mathcal{R}_P^{\mathrm{rsv}}\).
\(\mathcal{R}_u^{\mathrm{crb}}\) ranks candidate units using a scoring
dispatch \(\boldsymbol{P}^{\mathrm{score}}\). This dispatch starts as
\(\hat{\boldsymbol{P}}^{\mathrm{raw}}\) and is updated after each repair
cycle (Section~\ref{subsubsec:crb}).

The first cycle applies all operators. If the maximum overload still exceeds
\(\epsilon_{\mathrm{ov}}=10^{-3}\)~MW, a second cycle reruns
\(\mathcal{R}_u^{\mathrm{crb}}\), \(\mathcal{R}_u^{\mathrm{mud}}\), and
\(\mathcal{R}_P\). The procedure never solves a mixed-integer program (MIP).
Certification and fallback are described in Section~\ref{subsec:fallback}.

\subsubsection{Minimum Up/Down-Time Repair}
\(\mathcal{R}_u^{\mathrm{mud}}\) sweeps each generator chronologically while
tracking the remaining up- and down-time obligations inherited from the
pre-horizon state or created by repaired startups and shutdowns. A unit
with a pending up-obligation is kept online, a unit with a pending
down-obligation is kept offline, and otherwise the input state is
retained. It is applied before \(\mathcal{R}_u^{\mathrm{crb}}\) to enforce initial
obligations and after it to enforce obligations created by new startups in
Appendix~\ref{app:up_down_repair_details}.
 
\subsubsection{Capacity, Ramp, and Branch-Aware Commitment Repair}
\label{subsubsec:crb}
\(\mathcal{R}_u^{\mathrm{crb}}\) never deactivates an online unit. It starts only
eligible offline units, i.e., units whose minimum-down-time obligation
permits a startup in period \(t\), and extends each activation by the
minimum-up rule. For \(\zeta\in\{\mathrm{su},\mathrm{mc}\}\), let
\(C_{g,t}^{\zeta}\) denote the startup and marginal costs used for
ranking, which reduce to \(C_g^{\mathrm{su}}\) and \(C_g^{\mathrm{mc}}\)
in \eqref{eq:generator_feature_vector} when time-invariant. With
numerical tolerance \(\epsilon_{\mathrm{rep}}>0\), the normalized cost is
\begin{equation}
\widehat{C}_{g,t}^{\zeta}
=
\frac{C_{g,t}^{\zeta}}
{G^{-1}\sum_{j\in\mathcal{G}}C_{j,t}^{\zeta}+\epsilon_{\mathrm{rep}}},
\label{eq:normalized_cost}
\end{equation}
and, candidates are penalized by
\begin{equation}
\pi_{g,t}^{\mathrm{cost}}
=
\widehat{C}_{g,t}^{\mathrm{su}}
+
\widehat{C}_{g,t}^{\mathrm{mc}}.
\label{eq:common_cost_penalty}
\end{equation}
Candidates are evaluated with a scoring dispatch
\(\boldsymbol{P}^{\mathrm{score}}\), equal to
\(\hat{\boldsymbol{P}}^{\mathrm{raw}}\) in the first cycle and to the
preceding \(\mathcal{R}_P\) output in the repeated cycle, through the balanced
temporary dispatch
\begin{equation}
\boldsymbol{P}^{\mathrm{tmp}}
=
\mathcal{R}_P^{\mathrm{bal}}
\left(
\boldsymbol{P}^{\mathrm{score}}\odot\boldsymbol{u}^{\mathrm{work}}
\right),
\label{eq:temporary_balance_projection}
\end{equation}
where \(\boldsymbol{u}^{\mathrm{work}}\) is the current working
commitment, \(\odot\) is element-wise multiplication, and
\(\mathcal{R}_P^{\mathrm{bal}}\) in \eqref{eq:branch_repair_balance_correction}
uses the bounds induced by \(\boldsymbol{u}^{\mathrm{work}}\). Three
passes are applied in order.
 
\textit{Capacity pass:} For each period, offline candidates are
activated in descending order of
\(\hat{u}_{g,t}-\pi_{g,t}^{\mathrm{cost}}\) while
\(\sum_{g}P_g^{\max}u_{g,t}^{\mathrm{work}}<D_t\).
 
\textit{Ramp pass:} When ramp data are available, the same adequacy check
is repeated with each committed unit's output limited by what is reachable
from the neighboring periods of \(\boldsymbol{P}^{\mathrm{tmp}}\) under
\(RU_g\) and \(RD_g\). While this ramp-feasible
capability falls short of \(D_t\), the eligible candidate with the highest
score combining its ramp-feasible headroom, \(\hat{u}_{g,t}\), and
\(\pi_{g,t}^{\mathrm{cost}}\) is activated, one per period per iteration in
Appendix~\ref{app:commitment_repair_details}.
 
\textit{Branch pass:} Let \(\mathcal{L}_{\mathrm{act}}\subseteq\mathcal{L}\)
denote the active branches with finite positive limits. Under the
slack-bus convention in \eqref{eq:branch_repair_net_injection}, the
generator-to-branch sensitivity is
\begin{equation}
\Gamma_{\ell,g}
=
(\boldsymbol{H}\boldsymbol{A})_{\ell,g}-H_{\ell,s}.
\label{eq:branch_sensitivity_repair}
\end{equation}
Let \(\boldsymbol{F}^{\mathrm{tmp}}_t\) be obtained from
\eqref{eq:branch_repair_net_injection}--\eqref{eq:branch_repair_flow}
with \(\boldsymbol{P}^{\mathrm{bal},(k)}_t\) replaced by
\(\boldsymbol{P}^{\mathrm{tmp}}_t\). The most overloaded branch and the
directional sensitivity are
\begin{align}
\ell_t^{\mathrm{worst}}
&\in
\arg\max_{\ell\in\mathcal{L}_{\mathrm{act}}}
\left[|F_{\ell,t}^{\mathrm{tmp}}|-F_{\ell}^{\max}\right]_{+},
\label{eq:worst_branch_selection}\\
\kappa_{g,t}
&=
-\operatorname{sgn}\!\left(F_{\ell_t^{\mathrm{worst}},t}^{\mathrm{tmp}}\right)
\Gamma_{\ell_t^{\mathrm{worst}},g}.
\label{eq:directional_branch_sensitivity}
\end{align}
A positive \(\kappa_{g,t}\) means that increasing the output of
generator \(g\) moves the overloaded flow toward its limit. When the online units cannot provide enough relief
in Appendix~\ref{app:commitment_repair_details}, an offline candidate
with \(\kappa_{g,t}>\epsilon_{\mathrm{rep}}\) is activated according to
the highest score
\begin{equation}
s_{g,t}^{\mathrm{br}}
=
[\kappa_{g,t}]_{+}
\left[P_g^{\max}-P_g^{\min}\right]_{+}
+
0.01\hat{u}_{g,t}
-
\pi_{g,t}^{\mathrm{cost}}.
\label{eq:branch_repair_score}
\end{equation}
The score favors units with strong relief and large dispatch range,
uses the network prediction \(\hat{u}_{g,t}\) as a tie-breaker, and
penalizes expensive units. At most one unit per period is activated per
iteration; flows and scores are then recomputed until all overloads are
within \(\epsilon_{\mathrm{rep}}\), no candidate remains, or an
iteration cap is reached.
 
\subsubsection{Dispatch Repair}
With \(\boldsymbol{u}\) fixed, \(\mathcal{R}_P\) repairs dispatch in
three stages. First, it clips each output to
\([P_{g,t}^{\min}u_{g,t},P_{g,t}^{\max}u_{g,t}]\) and applies
\(\mathcal{R}_P^{\mathrm{bal}}\) to restore balance within these bounds. Second, it relieves overloaded active branches by PTDF redispatch. For a
branch \(\ell\) with flow \(F_{\ell,t}\), committed units are ranked by
\(-\operatorname{sgn}(F_{\ell,t})\Gamma_{\ell,g}\), so higher-ranked units
relieve the overload more. Output is shifted from a low-ranked donor to a
high-ranked receiver, within their headroom; because the shift is equal
and opposite, balance is preserved. Third, it corrects ramping, which the training-time repair ignores. Three
cycles alternate balance correction with forward and backward
chronological projection onto the ramp limits
\eqref{eq:ramp_up}--\eqref{eq:ramp_down}, followed by two ramp-aware
balance sweeps.

Because no further PTDF correction follows these sweeps,
branch overloads may remain or reappear.
Section~\ref{sec:numerical_experiments} reports the residual
violations. Tolerances and iteration caps are given in
Appendix~\ref{app:dispatch_repair_details}.

\subsubsection{Reserve Repair}
\label{subsubsec:reserve_repair}
Reserve shortfalls left after the first \(\mathcal{R}_P\) are repaired in
two stages. First, \(\mathcal{R}_u^{\mathrm{rsv}}\) only adds online hours. Visiting
shortfall periods chronologically, it greedily applies the startup or
online-block extension with the lowest added no-load and startup cost per
MW covered, subject to minimum up/down times, initial status, and
\(SU_g\), \(SD_g\). Then \(\mathcal{R}_u^{\mathrm{mud}}\) and
\(\mathcal{R}_P\) are reapplied. This repeats until the shortfall is
within \(\epsilon_{\mathrm{rsv}}\) or \(K_{\mathrm{rsv}}\) rounds are reached.

Second, \(\mathcal{R}_P^{\mathrm{rsv}}\) accounts for ramping, since
headroom alone overstates deliverable reserve when ramp limits bind. With
\(\tilde{\boldsymbol u}\) fixed, it shifts output within an hour from a
higher-cost donor to a receiver whose reserve in an adjacent hour is
ramp-limited. A move is kept only if it reduces the total shortfall and
keeps balance, bound, ramp, startup/shutdown, and branch limits satisfied.

\subsection{Feasibility Certification and Fallback}
\label{subsec:fallback}
The repaired schedule \((\tilde{\boldsymbol u},\tilde{\boldsymbol P})\)
is accepted when every constraint violation is within the tolerances of
Section~\ref{subsec:evaluation_metrics}. Otherwise, with
\(\tilde{\boldsymbol u}\) fixed, dispatch recovery reduces to a linear
program under the piecewise-linear costs in \eqref{eq:pwl_cost} (a convex
quadratic program for quadratic costs), which is solvable in polynomial
time, rather than a re-solution of the NP-hard NCUC MIP. Because the branch
limits in \eqref{eq:flow_limits} are softened by \(o_{\ell,t}\), this
program is infeasible only when \(\tilde{\boldsymbol u}\) lacks the
capacity, reserve, or ramping range to meet demand and
\eqref{eq:rsv_req}. In that case,
\(\tilde{\boldsymbol u}\) and \(\tilde{\boldsymbol P}\), when accepted by
the solver, would be supplied to the NCUC solver as a MIP start rather than as
a certified feasible incumbent. Note that this safeguard may not be exercised if FALCON-UC meets the tolerances.

\section{Numerical Experiments}
\label{sec:numerical_experiments}
\begin{table*}[t]
\centering
\caption{Performance of FALCON-UC on the 1354-bus system.
\(S_{\mathrm{mean}}\) is the mean end-to-end speedup over the solver.}
\label{tab:performance_1354}

\footnotesize
\renewcommand{\arraystretch}{1.15}
\setlength{\tabcolsep}{2pt}

\begin{tabular}{@{}lcccccccc@{}}
\toprule
System
& \(\mathrm{Acc}_{u}\)
& \(\lvert\Delta C\rvert_{\mathrm{mean}}\)
& \(e_{\mathrm{mean}}^{\mathrm{bal}}/e_{\max}^{\mathrm{bal}}\)
& \(V_{\mathrm{mean}}^{P}/V_{\max}^{P}\)
& \(V_{\mathrm{mean}}^{\mathrm{ramp}}/V_{\max}^{\mathrm{ramp}}\)
& \(V_{\mathrm{mean}}^{\mathrm{br}}/V_{\max}^{\mathrm{br}}\)
& \(\Delta_{\mathrm{mean}}^{\mathrm{rsv}}/\Delta_{\max}^{\mathrm{rsv}}\)
& \(S_{\mathrm{mean}}\) \\
& (\%)
& (\%)
& (MW)
& (MW)
& (MW)
& (MW)
& (MW)
& \\
\midrule
1354-bus
& 95.01
& 0.7176
& \(0.0015/0.0098\)
& \(0.0000/0.0000\) 
& \(0.0000/0.0000\) 
& \(0.0000/0.0000\)
& \(0.0000/0.0000\)
& \(1.89\times10^{3}\) \\
Solver \cite{santos2024unitcommitment} 
& --
& -- 
& \(0.0007/0.0037\) 
& \(0.0000/0.0000\) 
& \(0.0000/1.5{\times}10^{-5}\) & \(0.0000/0.0000\)
& \(0.0000/0.0000\)
& -- \\
\bottomrule
\end{tabular}

\vspace{1mm}
\end{table*}

\subsection{Experimental Setup}
\label{subsec:experimental_setup}

We compare the proposed method with three comparators:
\begin{enumerate}
    \item \textbf{Label-only baseline} (learning-only): a model with the
    same architecture, trained only with supervised commitment and
    dispatch losses. It is evaluated without correction or repair.
    \item \textbf{MTLR R-SCUC-FL}~\cite{ramesh2023feasibility}
    (solver-assisted): its predicted commitments are projected onto minimum up/down-consistent schedules and passed to a reduced NCUC solved with a MIP solver, with a confidence-based fallback if it is infeasible.
    \item \textbf{Solver}~\cite{santos2024unitcommitment}: the
    NCUC solved with UnitCommitment.jl.
\end{enumerate}

Augmented instances use the load profile
\begin{equation}
D^{\mathrm{aug}}_{b,t}
=
\max
\left\{
0,\;
D_{b,t}
\gamma
\left(1+\xi_t\right)
\left(1+\xi_b\right)
\right\},
\label{eq:load_augmentation}
\end{equation}
where \(\gamma\sim\mathcal{U}(0.90,1.10)\), \(\xi_t\sim\mathcal{N}(0,0.03^2)\), and \(\xi_b\sim\mathcal{N}(0,0.02^2)\). Here, \(\gamma\) is the global load-scaling factor, \(\xi_t\) is the time-dependent perturbation, and \(\xi_b\) is the bus-dependent perturbation; the maximum operator clips negative augmented loads to zero. For each system, the augmented dataset contains 668 independent 36-hour NCUC instances \((T=36)\), randomly divided into training, validation, and test subsets using an 80/10/10 split. The test set contains 66 instances, corresponding to a total of 2,376 operating hours.

\subsection{Evaluation Metrics}
\label{subsec:evaluation_metrics}

Let \(\mathcal{I}_{\mathrm{test}}=\{1,\ldots,N_{\mathrm{test}}\}\)
denote the test-sample index set. For FALCON-UC, all prediction,
economic, and feasibility metrics are computed from the final repaired
outputs \((\tilde{\boldsymbol{u}}_i,\tilde{\boldsymbol{P}}_i)\) and
the NCUC references
\((\boldsymbol{u}_i^*,\boldsymbol{P}_i^*)\). When the same metrics are
applied to the baseline or solver solutions, the corresponding
evaluated commitment and dispatch quantities are used. 

Commitment prediction accuracy and dispatch deviation are measured by
\begin{equation}
\mathrm{Acc}_{u}
=
\frac{100}{N_{\mathrm{test}}GT}
\sum_{i\in\mathcal{I}_{\mathrm{test}}}
\sum_{g\in\mathcal{G}}
\sum_{t\in\mathcal{T}}
\mathbb{I}
\left(
\tilde{u}_{i,g,t}
=
u_{i,g,t}^{*}
\right),
\label{eq:metric_commitment_accuracy}
\end{equation}
and
\begin{equation}
\mathrm{MAE}_{P}
=
\frac{1}{N_{\mathrm{test}}GT}
\sum_{i\in\mathcal{I}_{\mathrm{test}}}
\sum_{g\in\mathcal{G}}
\sum_{t\in\mathcal{T}}
\left|
\tilde{P}_{i,g,t}
-
P_{i,g,t}^{*}
\right|.
\label{eq:metric_pg_mae}
\end{equation}

Feasibility is evaluated using tolerances of \(\epsilon_P\),
\(\epsilon_{\mathrm{br}}\), \(\epsilon_{\mathrm{ramp}}\), and
\(\epsilon_{\mathrm{rsv}}\), each set to
\(0.001\)~MW, for the inequality constraints, and
\(\epsilon_{\mathrm{bal}}=0.01\)~MW for the power-balance equality constraint. For each constraint \(c\) below, \(x^{c}_{\mathrm{mean}}\) and
\(x^{c}_{\max}\) denote the mean and maximum of its error or violation
\(x^{c}\) over all corresponding test entries, and \(r^{c}_{\mathrm{fea}}\)
the percentage of entries within the associated tolerance.

For power balance,
\begin{equation}
e_{i,t}^{\mathrm{bal}}
=
\left|
\sum_{g\in\mathcal{G}}
\tilde{P}_{i,g,t}
-
\sum_{b\in\mathcal{B}}
D_{i,b,t}
\right|.
\label{eq:metric_balance_error}
\end{equation}

The commitment-dependent dispatch-bound violation is
\begin{equation}
V_{i,g,t}^{P}
=
\max
\left(
0,\,
P_{g,t}^{\min}\tilde{u}_{i,g,t}
-
\tilde{P}_{i,g,t},\,
\tilde{P}_{i,g,t}
-
P_{g,t}^{\max}\tilde{u}_{i,g,t}
\right).
\label{eq:metric_dispatch_bound_violation}
\end{equation}

For branch-flow feasibility, let
\(\widetilde F_{i,\ell,t}\) be computed from
\(\widetilde{\boldsymbol P}_{i}\) using the PTDF and slack-bus
convention in
\eqref{eq:branch_repair_net_injection}--\eqref{eq:branch_repair_flow}, with
\(\widetilde{\boldsymbol P}_{i}\) in place of \(\boldsymbol{P}^{\mathrm{bal},(k)}\).
The branch-flow violation is
\begin{equation}
V_{i,\ell,t}^{\mathrm{br}}
=
\left[
\left|\widetilde F_{i,\ell,t}\right|
-
F_{\ell}^{\max}
\right]_{+}.
\label{eq:branch_violation}
\end{equation} Maximum branch utilization is
\begin{equation}
U_{\max}^{\mathrm{br}}
=
100
\max_{i,\ell,t}
\frac{
\left|\widetilde F_{i,\ell,t}\right|
}{
F_{\ell}^{\max}
},
\label{eq:maximum_branch_utilization}
\end{equation}
excluding branches without a finite positive limit.
\begin{figure}[t]
\centering
\includegraphics[width=\columnwidth]{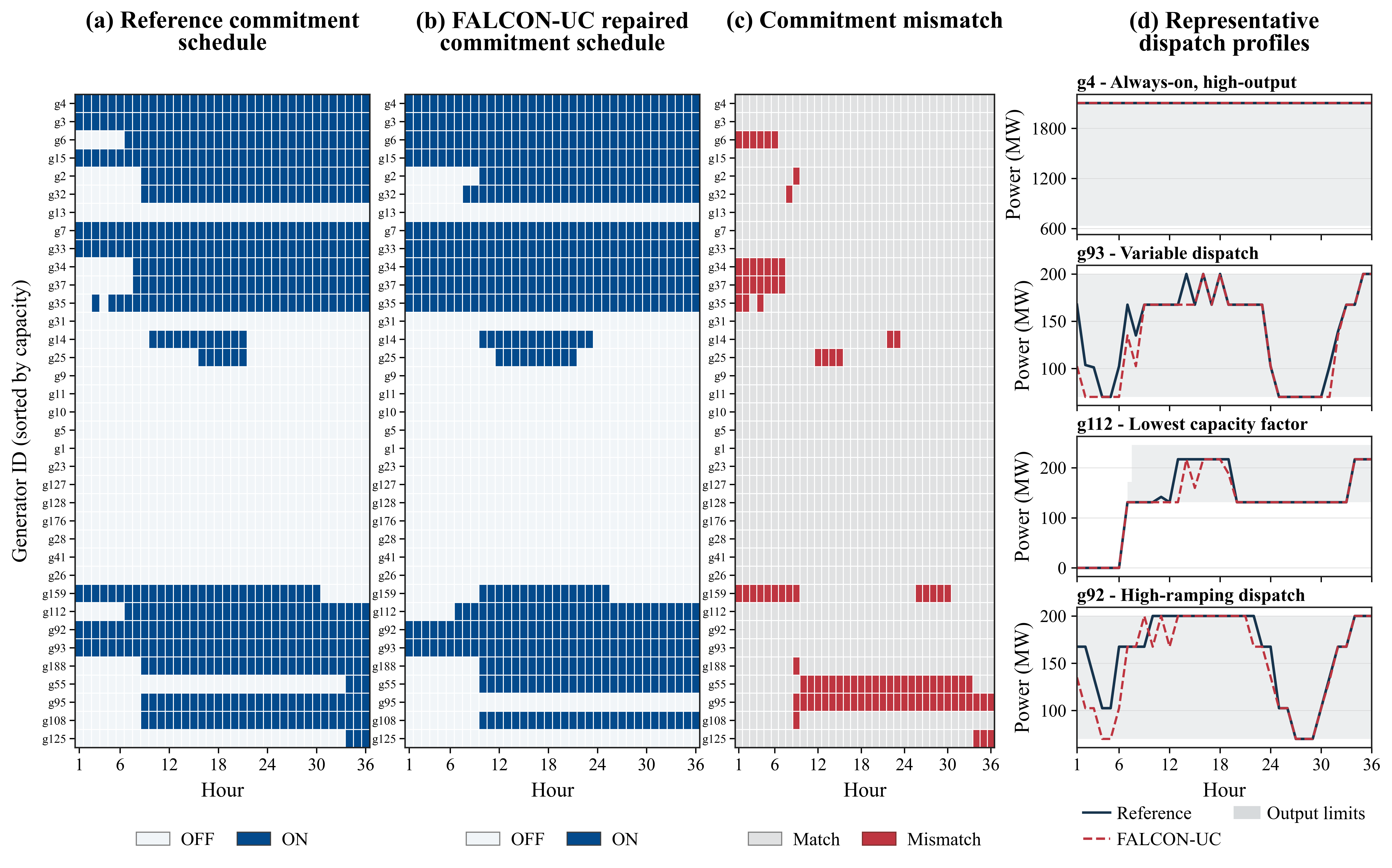}
\caption{Reference and FALCON-UC commitment and dispatch comparison on the 2383-bus system for a randomly selected sample from the test set.}
\label{fig:commitment_dispatch_2383}
\end{figure}

Ramping feasibility is evaluated using the NCUC ramping
constraints in \eqref{eq:ramp_up}--\eqref{eq:ramp_down} for all
\(i\in\mathcal{I}_{\mathrm{test}}\), \(g\in\mathcal{G}\), and
\(t=2,\ldots,T\). The repaired startup and shutdown indicators are
\begin{equation}
\tilde{v}_{i,g,t}
=
\left[
\tilde{u}_{i,g,t}
-
\tilde{u}_{i,g,t-1}
\right]_{+},
\qquad
\tilde{w}_{i,g,t}
=
\left[
\tilde{u}_{i,g,t-1}
-
\tilde{u}_{i,g,t}
\right]_{+},
\label{eq:repaired_startup_shutdown}
\end{equation}
where \(\tilde{u}_{i,g,0}=u_{g,0}\). The corresponding ramping
violation is
\begin{equation}
\begin{aligned}
V_{i,g,t}^{\mathrm{ramp}}
=
\max\Bigl\{
&0,\\
&\tilde{P}_{i,g,t}
-\tilde{P}_{i,g,t-1}
-RU_g\tilde{u}_{i,g,t-1}
-SU_g\tilde{v}_{i,g,t},\\
&\tilde{P}_{i,g,t-1}
-\tilde{P}_{i,g,t}
-RD_g\tilde{u}_{i,g,t}
-SD_g\tilde{w}_{i,g,t}
\Bigr\}.
\end{aligned}
\label{eq:metric_ramp_violation}
\end{equation}

Reserve adequacy is evaluated by the shortfall \(\Delta^{\mathrm{rsv}}_{i,t}\)
of the system reserve requirement \(R^{\mathrm{req}}_{i,t}\) against the
headroom that the online units of \(\tilde{\boldsymbol{u}}_i\) can still offer
above \(\tilde{\boldsymbol{P}}_i\), limited at startup and before shutdown.
Because this headroom ignores the ramp rows of the reference model, we also
report the shortfall \(\Delta^{\mathrm{rsv,ex}}_{i,t}\) against the ramp-coupled
reserve, in which each unit's headroom is further limited by its ramp-up and
ramp-down rows. The maximum and feasible rate of each, and the mean of
\(\Delta^{\mathrm{rsv}}\), are taken over all
\(i\in\mathcal{I}_{\mathrm{test}}\) and \(t\in\mathcal{T}\).

Let \(N_{i,\mathrm{up}}^{\mathrm{viol}}\) and
\(N_{i,\mathrm{down}}^{\mathrm{viol}}\) denote the numbers of
minimum up- and down-time violations of \eqref{eq:min_up} and
\eqref{eq:min_down} in sample \(i\), respectively. The reported totals
are
\[
N_{\mathrm{up}}^{\mathrm{viol}}
=
\sum_{i\in\mathcal{I}_{\mathrm{test}}}
N_{i,\mathrm{up}}^{\mathrm{viol}},
\qquad
N_{\mathrm{down}}^{\mathrm{viol}}
=
\sum_{i\in\mathcal{I}_{\mathrm{test}}}
N_{i,\mathrm{down}}^{\mathrm{viol}}.
\]
The metric \(r_{\mathrm{fea}}^{\mathrm{mud}}\) is the percentage of
test samples for which both counts are zero.

For each test sample \(i\), let
\(\tilde{C}^{\mathrm{tot}}_i\) and
\(C_i^{\mathrm{tot},*}\) denote the FALCON-UC and solver-reference
operating costs, respectively, using the definition in
Section~\ref{subsec:training_objective}. The percentage cost gap is
\begin{equation}
\Delta C_i
=
\frac{
\tilde{C}^{\mathrm{tot}}_i
-
C_i^{\mathrm{tot},*}
}{
|C_i^{\mathrm{tot},*}|+\epsilon_C
}
\times 100\%,
\label{eq:metric_cost_gap}
\end{equation}
where \(|\Delta C|_{\mathrm{mean}}\) denotes mean absolute cost gap.

\subsection{Numerical Results}
\label{subsec:numerical_results}
For the 1354-bus system, the maximum power-balance error of
\(0.0098\)~MW is within the \(0.01\)~MW tolerance, and all
2,376 balance entries remain feasible
in Table~\ref{tab:performance_1354}. No dispatch-bound, branch-flow, ramping, or minimum
up/down-time violations are observed, the maximum reserve shortfall is
\(0.0000\)~MW, and the inference-and-repair
time averages \(0.0517\)~s per sample.

On the 2383-bus system, Table~\ref{tab:prediction_cost} shows that
MTLR~R-SCUC-FL is closest to the solver reference on all three
metrics, with commitment accuracy \(99.48\%\) versus \(99.09\%\) for
FALCON-UC, dispatch MAE \(1.1150\)~MW versus \(1.6690\)~MW, and mean
absolute cost gap \(0.4155\%\) versus \(0.4350\%\). This advantage comes
from the reduced-NCUC solve that MTLR~R-SCUC-FL performs after
prediction, whereas FALCON-UC produces schedules without solving a MIP
(Table~\ref{tab:runtime}). The mean gaps are close, and FALCON-UC has
the smaller maximum gap (\(2.56\%\) versus \(5.61\%\)), with both
maxima occurring on the same two instances. Both methods substantially
reduce the baseline's dispatch MAE and cost gap (\(3.79\%\) versus
\(0.42\%\) and \(0.44\%\)); the baseline's commitment accuracy
(\(99.24\%\)) is slightly above FALCON-UC's but below MTLR~R-SCUC-FL's.
Fig.~\ref{fig:commitment_dispatch_2383} compares the reference and
FALCON-UC commitment and dispatch on the 2383-bus system.

\begin{table}[t]
\centering
\caption{Prediction and cost metrics on the 2383-bus system.}
\label{tab:prediction_cost}

\footnotesize
\renewcommand{\arraystretch}{1.25}
\setlength{\tabcolsep}{2.5pt}

\begin{tabular}{lccc}
\toprule
Metric & Baseline & \makecell[c]{MTLR\\R-SCUC-FL~\cite{ramesh2023feasibility}} & FALCON-UC \\
\midrule
\(\mathrm{Acc}_{u}\) (\%)
& 99.24 & \textbf{99.48} & 99.09 \\
\(\mathrm{MAE}_{P}\) (MW)
& 2.4159 & \textbf{1.1150} & 1.6690 \\
\(\lvert\Delta C\rvert_{\mathrm{mean}}\) (\%)
& 3.7917 & \textbf{0.4155} & 0.4350 \\
\(\lvert\Delta C\rvert_{\max}\) (\%)
& 12.7428 & 5.6071 & \textbf{2.5645} \\
\bottomrule
\end{tabular}
\end{table}

The feasibility results in Table~\ref{tab:feasibility} further
demonstrate that high commitment accuracy alone does not ensure an
operationally feasible solution. Despite its high commitment accuracy,
the baseline exhibits substantial power-balance, branch-flow, ramping,
and minimum up/down-time violations: its maximum balance error reaches
\(1374.9719\)~MW with no balance entry within tolerance, its worst branch
overload is \(1191.3024\)~MW, its worst
ramping violation is \(179.3386\)~MW, and only \(39.39\%\) of the test
instances satisfy the minimum up/down-time constraints. Its dispatch is
clamped to the commitment-dependent bounds, so it has no dispatch-bound
violations by construction.
In contrast, FALCON-UC eliminates all observed dispatch-bound, branch-flow,
ramping, minimum up/down-time, and spinning-reserve violations while
maintaining \(100\%\) power-balance feasibility under the prescribed
tolerances, with a maximum balance error of \(0.0024\)~MW, so it removes every
violation of the baseline. The headroom
repair is triggered in 7 of the 66 test instances and the ramp-coupled repair
in 2 of them, after which every test hour meets both the headroom and the
ramp-coupled requirement
(\(r^{\mathrm{rsv}}_{\mathrm{fea}}=r^{\mathrm{rsv,ex}}_{\mathrm{fea}}=100.00\%\)),
as do the MTLR~R-SCUC-FL and solver schedules.

MTLR~R-SCUC-FL and the solver reference~\cite{santos2024unitcommitment}
satisfy the power-balance, dispatch-bound, branch-flow, ramping, minimum
up/down-time, and reserve constraints, with maximum balance errors of
\(0.0013\)~MW and \(0.0012\)~MW. FALCON-UC, whose dispatch repair enforces
the PTDF branch limits as hard constraints, also keeps every branch within its
limit.

\begin{table}[t]
\centering
\vspace{1mm}
\caption{Feasibility metrics on the 2383-bus system.}
\label{tab:feasibility}

\scriptsize
\renewcommand{\arraystretch}{1.25}
\setlength{\tabcolsep}{2.5pt}

\begin{tabular}{@{}llcccc@{}}
\toprule
Constraint & Metric & Baseline &
\makecell[c]{MTLR\\R-SCUC-FL~\cite{ramesh2023feasibility}} &
FALCON-UC & Solver~\cite{santos2024unitcommitment} \\
\midrule

\multirow[c]{3}{*}{Power balance}
& \(e_{\mathrm{mean}}^{\mathrm{bal}}\) (MW)
& 296.2112 & \textbf{0.0002} & 0.0004 & 0.0002 \\
& \(e_{\max}^{\mathrm{bal}}\) (MW)
& 1374.9719 & \textbf{0.0013} & 0.0024 & 0.0012 \\
& \(r_{\mathrm{fea}}^{\mathrm{bal}}\) (\%)
& 0.00 & 100.00 & 100.00 & 100.00 \\
\midrule

\multirow[c]{3}{*}{Dispatch bounds}
& \(V_{\mathrm{mean}}^{P}\) (MW)
& 0.0000 & 0.0000 & 0.0000 & 0.0000 \\
& \(V_{\max}^{P}\) (MW)
& 0.0000 & 0.0000 & 0.0000 & 0.0000 \\
& \(r_{\mathrm{fea}}^{P}\) (\%)
& 100.00 & 100.00 & 100.00 & 100.00 \\
\midrule

\multirow[c]{4}{*}{Branch flow}
& \(V_{\mathrm{mean}}^{\mathrm{br}}\) (MW)
& 0.0487 & 0.0000 & 0.0000 & 0.0000 \\
& \(V_{\max}^{\mathrm{br}}\) (MW)
& 1191.3024 & 0.0006 & \textbf{0.0000} & 0.0006 \\
& \(r_{\mathrm{fea}}^{\mathrm{br}}\) (\%)
& 99.97 & 100.00 & 100.00 & 100.00 \\
& \(U_{\max}^{\mathrm{br}}\) (\%)
& 844.56 & 100.00 & 100.00 & 100.00 \\
\midrule

\multirow[c]{3}{*}{Ramping}
& \(V_{\mathrm{mean}}^{\mathrm{ramp}}\) (MW)
& 0.0214 & 0.0000 & 0.0000 & 0.0000 \\
& \(V_{\max}^{\mathrm{ramp}}\) (MW)
& 179.3386 & 0.0001 & \textbf{0.0000} & 0.0000 \\
& \(r_{\mathrm{fea}}^{\mathrm{ramp}}\) (\%)
& 99.94 & 100.00 & 100.00 & 100.00 \\
\midrule

\multirow[c]{4}{*}{\makecell[l]{Spinning\\reserve}}
& \(\Delta_{\max}^{\mathrm{rsv}}\) (MW)
& 0.0000 & 0.0000 & 0.0000 & 0.0000 \\
& \(r_{\mathrm{fea}}^{\mathrm{rsv}}\) (\%)
& 100.00 & 100.00 & 100.00 & 100.00 \\
& \(\Delta_{\max}^{\mathrm{rsv,ex}}\) (MW)
& 0.0000 & 0.0000 & 0.0000 & 0.0000 \\
& \(r_{\mathrm{fea}}^{\mathrm{rsv,ex}}\) (\%)
& 100.00 & 100.00 & 100.00 & 100.00 \\
\midrule

\multirow[c]{3}{*}{\makecell[l]{Minimum\\up/down time}}
& \(N_{\mathrm{up}}^{\mathrm{viol}}\)
& 66 & 0 & 0 & 0 \\
& \(N_{\mathrm{down}}^{\mathrm{viol}}\)
& 33 & 0 & 0 & 0 \\
& \(r_{\mathrm{fea}}^{\mathrm{mud}}\) (\%)
& 39.39 & 100.00 & 100.00 & 100.00 \\
\bottomrule
\end{tabular}

\vspace{1mm}
\parbox{\columnwidth}{\footnotesize
Values displayed as 0.0000 are below the reported numerical precision.}
\end{table}

Table~\ref{tab:runtime} summarizes the runtime performance. The
complete FALCON-UC inference-and-repair pipeline requires an average
of \(0.0562\)~s per sample and achieves a mean speedup of
\(2.27\times10^{3}\) relative to the solver. Although
MTLR~R-SCUC-FL has a similarly short NN forward pass, it needs a solver to complete each schedule, because its predictions are passed to a reduced-NCUC solve, which brings the average to \(262.2042\)~s per sample. In our implementation, this exceeds the solver's average, since infeasible fixed commitments trigger repeated reduced solves in its confidence-based fallback. By contrast, FALCON-UC
stays at the tens-of-milliseconds scale, mostly in its repair procedures.

\begin{table}[t]
\centering
\caption{Mean runtime and speedup on the 2383-bus system (solver and MTLR).}
\label{tab:runtime}

\small
\renewcommand{\arraystretch}{1.25}
\setlength{\tabcolsep}{3pt}

\begin{tabular}{@{}lccc@{}}
\toprule
Method
& \makecell[c]{Forward\\pass\\(s/sample)}
& \makecell[c]{End-to-end\\time\\(s/sample)}
& \makecell[c]{FALCON-UC\\speedup} \\
\midrule

FALCON-UC
& \textbf{0.0026}
& \textbf{0.0562}
& \(1.00\) \\

\makecell[l]{MTLR R-SCUC-FL\\[-1pt]
\cite{ramesh2023feasibility}}
& 0.0007
& 262.2042
& \(\mathbf{4.67\times10^{3}}\) \\

Solver~\cite{santos2024unitcommitment}
& --
& 127.8128
& \(\mathbf{2.27\times10^{3}}\) \\

\bottomrule
\end{tabular}

\vspace{1mm}
\end{table}

\section{Conclusion}
\label{sec:conclusion}
FALCON-UC combines LSTM--GraphSAGE commitment--dispatch prediction with differentiable training correction and deterministic inference repair, including spinning-reserve repair, without solving a MIP at inference time. On the 2383-bus system, all evaluated constraints, including ramp-coupled spinning reserve, remain within tolerance at \(0.0562\)~s per sample, yielding a \(2.27\times10^{3}\) speedup over the solver. MTLR~R-SCUC-FL has higher commitment accuracy (\(99.48\%\) versus \(99.09\%\)), lower dispatch error, and a slightly lower cost gap (\(0.4155\%\) versus \(0.4350\%\)), but it depends on a reduced-NCUC solve that takes \(262.2\)~s per instance, \(4.67\times10^{3}\) times longer than FALCON-UC. Every repaired test schedule met the stated tolerances, so the certification-and-fallback step in Section~\ref{subsec:fallback} was not used.

Evaluation is limited to load perturbations around base instances, so generalization to structurally different conditions is unestablished. The repair is heuristic and gives no exact feasibility or optimality guarantee. The experiments do not isolate the correction from the repair. Future work will address topology-informed generator graphs, stronger repair guarantees, renewable uncertainty, \(N-1\) contingencies, broader distributions, and transfer to unseen systems.

\appendices
 
\section{Minimum Up/Down-Time Repair}
\label{app:up_down_repair_details}
 
For a binary input schedule \(\boldsymbol{u}^{\mathrm{in}}\), let
\(\boldsymbol{u}^{\mathrm{out}}=\mathcal{R}_u^{\mathrm{mud}}(\boldsymbol{u}^{\mathrm{in}})\)
and \(d_{g,t}=\mathbb{I}(u_{g,t}^{\mathrm{in}}\geq0.5)\). Let
\(I_g>0\) denote the number of consecutive pre-horizon periods for which
generator \(g\) has been online, and \(-I_g>0\) the number for which it
has been offline. The remaining up- and down-time obligations
\(q_{g,t}^{\mathrm{up}}\) and \(q_{g,t}^{\mathrm{down}}\) at the
beginning of period \(t\) are initialized as
\begin{equation}
\begin{aligned}
q_{g,1}^{\mathrm{up}}
&=
\begin{cases}
[T_g^{\mathrm{up}}-I_g]_{+},&I_g>0,\\
0,&I_g\leq0,
\end{cases}
\\
q_{g,1}^{\mathrm{down}}
&=
\begin{cases}
[T_g^{\mathrm{down}}+I_g]_{+},&I_g<0,\\
0,&I_g\geq0,
\end{cases}
\end{aligned}
\label{eq:app_mud_counter_initialization}
\end{equation}
or to zero when initial-status data are unavailable. The repaired state
is selected chronologically as
\begin{equation}
u_{g,t}^{\mathrm{out}}
=
\begin{cases}
1,
& q_{g,t}^{\mathrm{up}}>0,\\
0,
& q_{g,t}^{\mathrm{up}}=0
  \ \text{and}\ q_{g,t}^{\mathrm{down}}>0,\\
d_{g,t},
& \text{otherwise}.
\end{cases}
\label{eq:app_mud_repaired_state}
\end{equation}
Period~1 is a genuine transition from the pre-horizon state
\(u_{g,0}=\mathbb{I}(I_g>0)\), so for \(t=1,\ldots,T\) the repaired startup and
shutdown indicators are
\(v_{g,t}^{\mathrm{out}}=[u_{g,t}^{\mathrm{out}}-u_{g,t-1}^{\mathrm{out}}]_{+}\)
and
\(w_{g,t}^{\mathrm{out}}=[u_{g,t-1}^{\mathrm{out}}-u_{g,t}^{\mathrm{out}}]_{+}\)
with \(u_{g,0}^{\mathrm{out}}=u_{g,0}\), and for \(t=1,\ldots,T-1\),
\begin{align}
q_{g,t+1}^{\mathrm{up}}
&=
\begin{cases}
T_g^{\mathrm{up}}-1,
& v_{g,t}^{\mathrm{out}}=1,\\
\left[q_{g,t}^{\mathrm{up}}-1\right]_{+},
& \text{otherwise},
\end{cases}
\label{eq:app_mud_up_counter_update}
\\
q_{g,t+1}^{\mathrm{down}}
&=
\begin{cases}
T_g^{\mathrm{down}}-1,
& w_{g,t}^{\mathrm{out}}=1,\\
\left[q_{g,t}^{\mathrm{down}}-1\right]_{+},
& \text{otherwise}.
\end{cases}
\label{eq:app_mud_down_counter_update}
\end{align}
A startup (shutdown) in period~1 thus keeps the generator online (offline)
through period \(T_g^{\mathrm{up}}\) (\(T_g^{\mathrm{down}}\)), as for later
transitions; without initial-status data, no period-1 transition is defined. For
\(T=1\), only
\eqref{eq:app_mud_counter_initialization}--\eqref{eq:app_mud_repaired_state}
apply. A generator that is online before the horizon with
\(P_{g,0}>SD_g\) is held online for \(\lceil(P_{g,0}-SD_g)/RD_g\rceil\) periods
(the whole horizon if \(P^{\min}_g>SD_g\)) before
\eqref{eq:app_mud_repaired_state} is applied, and an off gap shorter than
\(T_g^{\mathrm{down}}\) after them is filled.

\section{Commitment-Repair Details}
\label{app:commitment_repair_details}
 
In the capacity pass, candidates are examined once per period;
capacity and startup permission are recomputed after each activation,
but the ranking is not.
 
\textit{Ramp pass:} Each iteration recomputes
\(\boldsymbol{P}^{\mathrm{tmp}}\) by
\eqref{eq:temporary_balance_projection} and defines the neighbor bounds
\(B_{g,t}^{-}=P_{g,t-1}^{\mathrm{tmp}}+RU_g\) for \(t>1\)
and \(B_{g,t}^{+}=P_{g,t+1}^{\mathrm{tmp}}+RD_g\) for
\(t<T\), with inactive bounds set to \(+\infty\). The committed
ramp-feasible capability and offline-candidate headroom are
\begin{align}
U_{g,t}
&=
\max
\left\{
P_g^{\min}u_{g,t}^{\mathrm{work}},
\min
\left(
P_g^{\max}u_{g,t}^{\mathrm{work}},
B_{g,t}^{-},
B_{g,t}^{+}
\right)
\right\},
\nonumber\\
\Delta P_{g,t}^{\mathrm{cand}}
&=
\left[
\min\!\left(P_g^{\max},B_{g,t}^{-},B_{g,t}^{+}\right)
-P_g^{\min}
\right]_{+}.
\label{eq:app_ramp_capability_headroom}
\end{align}
A shortage exists when
\(\sum_gU_{g,t}<D_t-\epsilon_{\mathrm{rep}}\). An offline,
startup-permitted unit with
\(\Delta P_{g,t}^{\mathrm{cand}}>\epsilon_{\mathrm{rep}}\) is then
eligible and is ranked by
\(\Delta P_{g,t}^{\mathrm{cand}}+0.01\hat{u}_{g,t}-\pi_{g,t}^{\mathrm{cost}}\).
The pass stops when no shortage remains, no candidate exists, or the
iteration cap is reached.
 
\textit{Branch pass:} Let
\(\kappa_t^{\mathrm{down}}=\min(\min_{g}\kappa_{g,t},0)\),
\(\Delta\kappa_{g,t}=[\kappa_{g,t}-\kappa_t^{\mathrm{down}}]_{+}\), and
\(\Delta P_{g,t}^{\mathrm{br,up}}
=[P_g^{\max}u_{g,t}^{\mathrm{work}}-P_{g,t}^{\mathrm{tmp}}]_{+}\).
The relief from raising online units against the most effective donor is
\(\Psi_t^{\mathrm{pair}}=\sum_{g}\Delta P_{g,t}^{\mathrm{br,up}}\Delta\kappa_{g,t}\),
and the relief from raising units with \(\kappa_{g,t}>0\) alone is
\(\Psi_t^{\mathrm{dir}}=\sum_{g}\Delta P_{g,t}^{\mathrm{br,up}}[\kappa_{g,t}]_{+}\).
An additional unit is requested when the worst
violation exceeds \(\epsilon_{\mathrm{rep}}\) and
\(\Psi_t^{\mathrm{pair}}\) is smaller than that violation plus
\(\epsilon_{\mathrm{rep}}\). Optionally, a unit is requested only if
\(\Psi_t^{\mathrm{dir}}\) is also insufficient.

\section{Dispatch-Repair Details}
\label{app:dispatch_repair_details}
 
The balance step in \(\mathcal{R}_P\) stabilizes the headroom denominator in
\eqref{eq:branch_repair_balance_weight} by \(10^{-6}\). Each PTDF transfer
includes a \(10^{-3}\)~MW overload margin, and the PTDF pass stops when the
maximum overload is at most \(10^{-6}\)~MW, its improvement is at most
\(10^{-4}\)~MW, or after 15 iterations (10 in the second application in
Algorithm~\ref{alg:inference_repair}). In the balance--ramp cycles, an empty ramp
interval retains the generator-bound projection, and no
pre-horizon-to-first-period ramp projection is performed.

\renewcommand{\baselinestretch}{0.85} 
\bibliographystyle{IEEEbib} 
\bibliography{references-of-work}

\end{document}